# Topology-Optimized Dielectric Cavities for Enhanced Excitonic Light Emission from WSe₂

Owen Matthiessen,[1, 2, *] Brandon Triplett,[1, 2, *] Omer Yesilyurt,[1, 2, *] Davide Cassara,[3, *] Karthik Pagadala,[1, 2]
Morris M. Yang,[1, 2] Andres E. Llacsahuanga Allcca,[4] Hamza Ather,[4, 1, 2] Colton Fruhling,[1, 2] Abhishek
Bharatbhai Solanki,[1, 2] Yong P. Chen,[1, 5, 2, 4] Hadiseh Alaeian,[1, 5, 2, 4] Alexander V. Kildishev,[1, 2] Vladimir
M. Shalaev,[1, 5, 2, 4, †] Federico Capasso,[3, ‡] Alexandra Boltasseva,[1, 5, 2, §] and Vahagn Mkhitaryan[1, 2, ¶]

[1]*Elmore Family School of Electrical and Computer Engineering,*
*Purdue University, West Lafayette, IN 47906, USA*
[2]*Birck Nanotechnology Center, Purdue University, West Lafayette, IN 47906, USA*
[3]*John A. Paulson School of Engineering and Applied Sciences,*
*Harvard University, Cambridge, MA 02138, USA*
[4]*Department of Physics and Astronomy, Purdue University, West Lafayette, IN 47907, USA*
[5]*Purdue Quantum Science and Engineering Institute,*
*Purdue University, West Lafayette, IN 47907, USA*
(Dated: September 30, 2025)

Photonic inverse design and, especially, topology optimization, enable dielectric cavities with deeply sub-diffraction mode volumes and high quality factors, thus offering a powerful platform for enhanced light-matter coupling. Here, we design and fabricate arrays of CMOS-compatible silicon cavities on sapphire with extreme subwavelength transverse mode sizes of only 30-40 nm (V ∼ λ³/2500). These cavities are engineered for deterministic coupling to a monolayer (or few-layer) excitonic material, producing strong near-field localization directly beneath the 2D material. Photoluminescence (PL) measurements show reproducible tenfold enhancements relative to bare silicon, consistent with numerical simulations that account for material absorption and fabrication tolerances. Furthermore, time-resolved PL measurements reveal pronounced lifetime shortening and non-exponential dynamics, indicating cavity-mediated exciton-exciton interactions. The optimized cavity geometry enhances the far-field collection efficiency and supports scalable integration with van der Waals semiconductors. Our results show that the arrays of topology-optimized dielectric cavities are a versatile, scalable platform for controlling excitonic emission and interactions, which creates new opportunities in nonlinear optics, optoelectronics, and quantum photonics.

## I. INTRODUCTION

Strong and tailorable light-matter interaction at the nanoscale is essential for modern photonics. It enables on-chip quantum and nonlinear optics, low-threshold lasing, and efficient photon sources. Achieving this regime requires confining electromagnetic energy efficiently in modes with small mode volumes and high quality factors (Q-factors). Plasmonic nanostructures have long enabled extreme subwavelength confinement and Purcell enhancement [1, 2], offering the potential to greatly accelerate radiative processes [3, 4] and overcome decoherence [5] in semiconductor quantum dots and other solid state and molecular emitters [2, 4] as well as 2D materials [6–16]. However, they are fundamentally limited by ohmic losses. In quantum applications, these losses can cause photons generated by quantum emitters to be reabsorbed by plasmons, reducing efficiency and coherence. All-dielectric nanophotonic cavities provide a compelling alternative [17–24]. They combine low loss with com-patibility for large-scale photonic integration. The performance of conventional dielectric cavities, however, is often constrained by a large mode volume. In such traditional architectures, such as photonic crystals [25–28], microring resonators [29–32], and Distributed Bragg Reflector (DBR) cavities [33–36], this constraint restricts field localization to the diffraction limited volume and consequently results in weaker emitter-cavity coupling.

Recent advances in photonic inverse design and topology optimization have led to a new class of dielectric and plasmonic cavities that support optical modes with extreme field enhancement and compact mode volumes [37–47]. For dielectric structures, this approach exploits the full design freedom of computational optimization to concentrate light into deep-subwavelength regions, achieving mode volumes comparable to plasmonic systems but with negligible absorption losses. Initially applied to optimize grating couplers[42, 48, 49], topology optimization has since evolved to incorporate fabrication constraints in the inverse design of plasmonic and dielectric nanophotonic devices [38, 39, 50, 51]. Early implementations in plasmonic structures have improved field confinement and directional emission from quantum emitters [41, 52, 53], while more recent efforts have shifted toward low-loss dielectric platforms; using silicon [54], group III-V materials such as gallium (GaP)[40] and indium phosphide (InP)[55]. These advances have enabled experimental demonstrations in nanolasing [45, 46], enhanced fluorescence [41], strong coupling with monolayer transition metal dichalcogenides (TMDC) at cryogenic temperatures [47], nonlinear optical enhancement [55], and efficient light-matter interactions [56, 57].

---

* These authors contributed equally.
† Corresponding author: shalaev@purdue.edu
‡ Corresponding author: capasso@seas.harvard.edu
§ Corresponding author: aeb@purdue.edu
¶ Corresponding author: vmkhitar@purdue.edu

Two-dimensional van der Waals (2D-vdW) semiconductors, such as $WSe_2$ and other TMDCs, offer exciting opportunities for photonics and quantum photonics applications. They host tightly bound excitons with large oscillator strengths and are compatible with a wide range of substrates. These materials are ideal platforms for exploring rich exciton and exciton-polariton physics [58]. Applications range from photodetection [59, 60] and light emission [29, 61] to single-photon-level nonlinearities [47, 62], valleytronics [63, 64], and solid-state quantum simulation [65, 66]. In recent years, they have been the focus of intense research on various types of excitons [67, 68] and exciton complexes [69–72] in both monolayers [70, 73] and Moiré heterostructures [65, 66, 74–76]. However, their atomic thickness limits the volume for interaction with free-space optical fields, requiring engineered photonic environments to boost light-matter coupling.

In this work, we explore the coupling of a $WSe_2-$monolayer to 2D-periodic arrays of identical topology-optimized silicon dielectric cavities fabricated on sapphire substrates. A deeply subwavelength gap in the center of each inverse-designed cavity enables strong localization of optical fields directly beneath the monolayer. This approach provides a straightforward way to create deterministic hotspots while retaining control over nanometer-scale light confinement. Such capability is advantageous compared to plasmonic gap modes, which are widely regarded as among the best platforms for achieving extreme field enhancements but offer limited deterministic emitter-hotspot alignment due to the difficulty of controllably placing chemically synthesized nanoparticles.

Numerical simulations of the studied topology-optimized cavity arrays reveal strong optical field confinement in nanoscale hotspots at the center of each cavity, with near-field amplitude enhancements exceeding $50\times$ relative to the incident field under realistic material losses in silicon and absorption by the $WSe_2$ monolayer (for detailed discussion, see II Section). Additionally, the dipole emission simulations show a marked increase in the emission directionality when the emitter is aligned with the linearly polarized bowtie-like mode of the cavity, compared to emission from $WSe_2$ on flat, unpatterned silicon.

Experimentally, we observe a consistent photoluminescence enhancement of approximately $10\times$ when comparing off-cavity regions, confirmed both by spatial PL maps and individual spectra. , dark excitons, and biexcitons seen previously in other studies [70, 77–83](see IV Fig. S1). With the metasurface cavities, these features are better defined than on flat regions of the plane substrate, with narrower linewidths. This behavior is consistent with cavity-assisted Purcell enhancement of the radiative decay rates and diminished contribution from non-radiative pathways.

Furthermore, time-resolved PL measurements reveal that the exciton decay dynamics become substantially non-exponential in the presence of the cavity in the entire range of pump powers. In contrast, the decay remains nearly exponential over a broad range of pump powers in the absence of the cavity, deviating only at the higher powers. The observed difference between on- and off-cavity behavior, together with its power dependence, points to the onset of multi-exciton interactions and cavity-enhanced energy-transfer processes [84–88].

These observations highlight the potential of topology-optimized dielectric metasurface cavities for controlling emission processes and excitonic interactions in atomically thin semiconductors, and for probing nonlinear and many-body phenomena in strongly confined optical environments.

## II. RESULTS

In this study, we use photoluminescence (PL) spectroscopy to investigate the interaction of excitons in $WSe_2$ monolayers (MLs) and topology-optimized dielectric cavities with subwavelength mode confinement. The cavities were developed by using the photonic inverse design approach, which is described in detail in the Methods section. The metasurface cavities consist of a 100 - 200 nm-thick Si cavity layer on top of a sapphire substrate. Although silicon is relatively lossy at the exciton emission wavelengths in $WSe_2$, it was chosen for its availability, technological relevance, and ease of fabrication. These intrinsic losses limit the achievable quality factor; however, the high refractive index of Si enables strong field confinement within small spatial regions and allows us to achieve strong interaction between this mode and excitons. While silicon is absorptive at the operating wavelengths, it is essentially lossless in the near-infrared and is an ideal platform for such cavity applications[43]. Additionally, in our design, we deliberately used supported cavities rather than suspended membranes, as this configuration offers greater mechanical robustness and practicality for scalable integration of 2D materials, in contrast to suspended photonic devices.

Large-area (typically $\sim 40 \times 30 \mu m^2$) hBN-capped $WSe_2$ monolayers were transferred onto the cavity arrays using a gold-assisted transfer method [89]. Each flake was carefully positioned to lie on both the cavity array and on the adjacent flat Si-on-sapphire substrate, as shown in Fig. 1a, allowing for a direct on-chip comparison between cavity-coupled and reference (uncoupled) emission.In the following, we refer to regions of the $WSe_2$ flake located on top of the cavity arrays as 'on-cavity', while the regions located on the adjacent unpatterned Si/sapphire substrate are referred to as 'off-cavity'. After the transfer, the strain on the flakes, which is in-duced by proximity to the cavity was characterized using Raman spectroscopy. We found that the strain inhomogeneity is negligible, with Raman peak shifts of less than 1 cm⁻¹ compared to the reference region of the flake on the flat Si/sapphire substrate. A detailed discussion of the Raman maps and shift analysis is provided in IV Fig. S2. Figure 1b shows an example optical microscope image of an hBN/WSe₂ flake positioned over one of the studied cavity arrays studied in our experiments. The image includes the PL map from Fig. 1c overlaid to indicate the location of the WSe₂ flake beneath the hBN encapsulation. The PL map in Fig. 1c clearly reveals the enhanced emission from the cavity region compared with the weaker response observed from the off-cavity reference. We studied three different topology-optimized metasurface cavities (Fig. 1 panels d-o) that were obtained through the same optimization process but with different random initiations of the algorithm, with the importanat exception that the cavity in Fig. 1l which was optimized for the case when WSe₂ is present. Our optimization process consistently converges to a designs featuring a bowtie-like structure in the middle of the unit cell as it is shown in SEM images in Fig. 1d, h and l with the red outlines. Such consistent appearance of the bowties in all devices indicates the presence of a global minimum with this feature, while variations in the surrounding dielectric distribution correspond to nearby local minima to which algorithm converges.

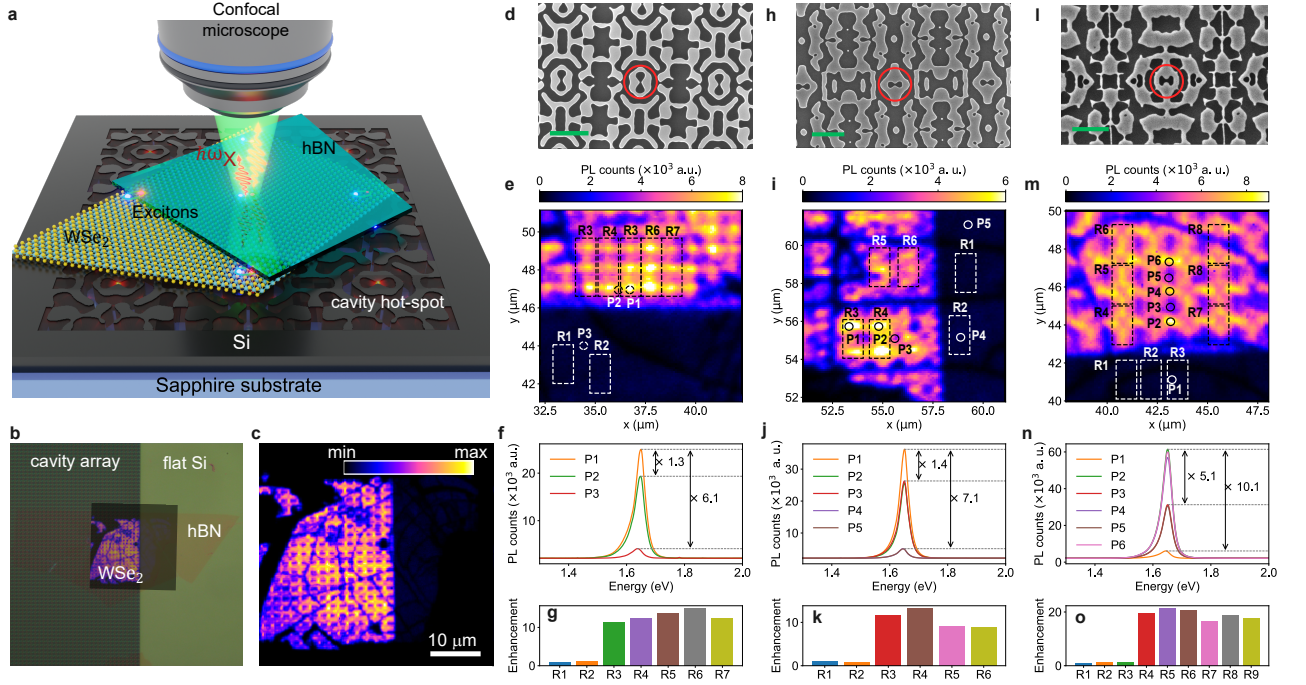

Figure 1. **PL maps and enhancement of WSe₂ coupled to topology-optimized cavities. a**, Illustration of the confocal microscopy setup used to collect PL maps from the hBN/WSe₂/topology-optimized cavity/sapphire stack. In all experiments, the flakes are positioned such that part of the WSe₂ flake lies on the cavity array and part on the unpatterned Si region, providing a reference signal. **b,c**, Optical microscope image at the edge of a cavity array, with the inset PL map superimposed to indicate the location of the WSe₂ flake relative to the cavity boundary since PL emission occur predominantly from WSe₂. The PL map is collected over a $40 \times 40 \ \mu m^2$ region, covering both cavity and reference areas. **d-o**, Data for three representative samples, including SEM images, high-resolution PL maps, PL intensity counts, and enhancement histograms. (**e**),(**i**),and(**m**) show the PL maps of all three samples. The circular markers indicate the positions where PL spectra shown in (**f**),(**j**),and(**n**) were measured. The dashed rectangles denote the regions used to compute the average PL enhancement factors shown in (**g**),(**k**),and(**o**). The identical rectangular regions were selected on the flat Si reference and the patterned cavity arrays to ensure the consistent comparison. The scale bars on SEM images in (**d**), (**h**) and (**l**) are all 500 nm.

The panels in Fig. 1d-f show scanning electron microscopy (SEM) images (Fig. 1d), high-resolution PL map (Fig. 1e), integrated PL spectra over 10 s (Fig. 1f), and the histograms of PL enhancement factors (Fig. 1g) of our first cavity. The circular markers in Fig. 1e indicate the locations of the points for which the spectra are shown in panel Fig. 1f. The arrows in panel Fig. 1f highlight the relative enhancement of peak counts for different positions on the cavity. For this specific cavity in particular, we see more than 6 times enhancement of the PL peak counts compared to the off-cavity region was observed. Figure 1g shows the average PL enhancement, obtained by comparing the mean PL counts from identically sized rectangular regions on and off the cavity arrays in the PL maps of Fig. 1e. For this sample, the analysis yields the enhancement of approximately 14 times. Fig-ure 1h-k and Fig. 1l-o present the equivalent measurements and analyses for the two other cavity arrays with different geometries. Among those, the cavity shown in Fig. 1l-o exhibits the largest enhancement- nearly twice of what is observed for the other two arrays. We want to emphasize that the enhancement factors in Fig. 1f,j,n are calculated with respect to point $P1$ which is in the off-cavity region for all the samples. These differences can be partly explained by the resonance shifts relative to the exciton emission wavelength, arising from fabrication tolerances, as well as variations in the observed collection efficiency, since some cavities provide superior emission directionality compared to others (see the discussion below). The combined effect of these factors leads to the observed variations in enhancement across different cavities in our experiments.

We performed numerical simulations for the designed cavities shown in Fig. 1. These include studies of the near-field distributions, far-field radiation patterns, and the resonance spectra (Fig. 2). The simulated near-field distributions of a cavity array unit cell overlaid on the scanning electron microscopy (SEM) images (Fig. 2a,b) clearly revealing the strong localization of the cavity mode within the bowtie-like gap region at the cavity center. Figure 2c-e display the far-field radiation patterns (left panel) and the corresponding directivity plots (right panel) for a dipole emitter placed on the off-cavity region (Fig. 2c) and at the center of bow-tie like regions of two representative cavities in Fig. 2d and Fig. 2e. In all cases, the dipole is aligned with the cavity mode polarization and positioned at the top surface plane of the structure. The simulations incorporate the full material stack, including the $WSe_2$ monolayer and the encapsulating hBN layer, with the precise hBN thickness values determined from atomic force microscopy (AFM) measurements (see IV Fig. S3). The dipole radiation pattern plots (Fig. 2c-e) reveal that, a dipole placed on the off-cavity region primarily emits at grazing angles relative to the surface normal. However, a dipole of the same polarization placed in the 'hot-spot' region at the top of the cavity surface exhibits highly directional emission. The significant fraction of the radiated power for this case is confined within the numerical aperture of our collection objective (NA = 0.9). The radiation directivity $D(\theta, \phi)$ in Fig. 2c-e is defined as $D(\theta, \phi) = 4\pi I(\theta, \phi)/P_{tot}$, where $I(\theta, \phi)$ is the radiated intensity in direction $(\theta, \phi)$ and $P_{tot}$ is the total radiated power. It quantifies how directional the emission is compared to an isotropic radiator $(D = 1)$.

The near-field distributions were simulated for two cross-sections (Fig. 2f and g): the x-y plane at the position of the $WSe_2$ monolayer (Fig. 2f), and the x-z plane across the cavity at $y = 0$ (Fig. 2g) for the cavity shown in Fig. 1l. These near-field distributions are extracted at the exciton emission energy of 1.65 eV. For the ideal cavity geometry, the simulations predict up to a 50-fold enhancement of the local field intensity. Figure 2h and i show the near-field distributions for the same cavity design but with the contours extracted from the experimental SEM images (see IV Fig. S4). In our experiments, however, the small feature sizes and high pattern density increase the proximity effects in electron beam lithography (EBL) causing slight shape distortions from the original design. Combined with the roughness introduced during reactive-ion etching, these deviations increase radiative losses, reducing the field localization, and limit the attainable enhancement. These effects are especially pronounced for cavity resonances at shorter wavelengths, where required feature sizes shrink and fabrication tolerances tighten proportionally. We show that the near-field intensity enhancement is reduced to approximately 10-fold (Fig. 2h), in agreement with the experimentally observed photoluminescence enhancement and lifetime shortening. This reduction arises not only from the geometric imperfections that introduce light scattering and cause destructive interference, but also from the resonance shifts and broadening resulted . Indeed, Fig. 2j shows the simulated absorption spectra for the ideal (solid lines) and SEM-extracted (dashed lines) cavity geometries, both with and without the $WSe_2$/hBN stack. The presence of $WSe_2$ introduces additional resonance broadening and redshifts due to absorption and radiative losses. Fabrication imperfections further exacerbate these effects, resulting in significant degradation of the field enhancement. A similar trend is evident in the calculated radiative decay rate ratios on and off the cavity (Fig. 2k).

We further investigated the exciton decay dynamics for the on- and off-cavity cases using time-resolved photoluminescence (TRPL) measurements. These measurements were carried out using a femtosecond, pulsed ($< 100$ fs, 80 MHz repetition rate) laser light with $\lambda = 520$ nm central wavelength, providing energy above the $WSe_2$ band gap. Figure 3a shows PL maps with open circles indicating the locations where decay curves in Fig. 3b,c were recorded. Fig. 3b,c display the decay curves for on and off the cavity cases, respectively, measured at three different pump powers: 2.5, 5, and 15 $\mu$W. These measurements reveal clear non-exponential decay dynamics. This effect is most pronounced on the cavities, where the non-exponential decays are observed at all measured pump powers (see Fig. 3b). In contrast, for off-cavity regions, such behavior appears only at higher pump powers (Fig. 3c). In both on- and off-cavity cases, we observe a clear shortening of the decay time with increasing pump power, accompanied by a more pronounced non-exponential character. This behavior suggests that the decay dynamics are more likely governed by exciton–exciton interactions, as higher pump powers generate larger exciton populations, increase their density, and thereby reduce inter-exciton distances in the pump region. Similar non-exponential decay signatures at elevated pump powers have been reported in earlier studies of $WSe_2$ on unpatterned substrates, although their microscopic origin remains debated [90–92]. Depending on the experimental conditions and material qual-ity, this type of dynamics have been attributed to a range of mechanisms, including exciton-exciton interactions [90], defect-assisted trapping and tunneling [92], and population transfer between bright and dark exciton states [91]. We believe that in our samples both defect-related processes and exciton–exciton interactions contribute to the observed non-exponential decay. Similar to Ref. [92], we observe a pronounced long-time tail in the decay curves, which can be attributed to the localization effects and inhomogeneous broadening. The presence of such defects is further supported by our low-temperature spectra, where distinct low-energy emission features consistent with the defect-bound excitons are observed. At the same time, the strong power dependence of the non-exponential dynamics provides a compelling evidence that exciton–exciton interactions play a dominant role, particularly in the cavity regions where the local fields significantly enhance resonant dipole-dipole interaction effects.

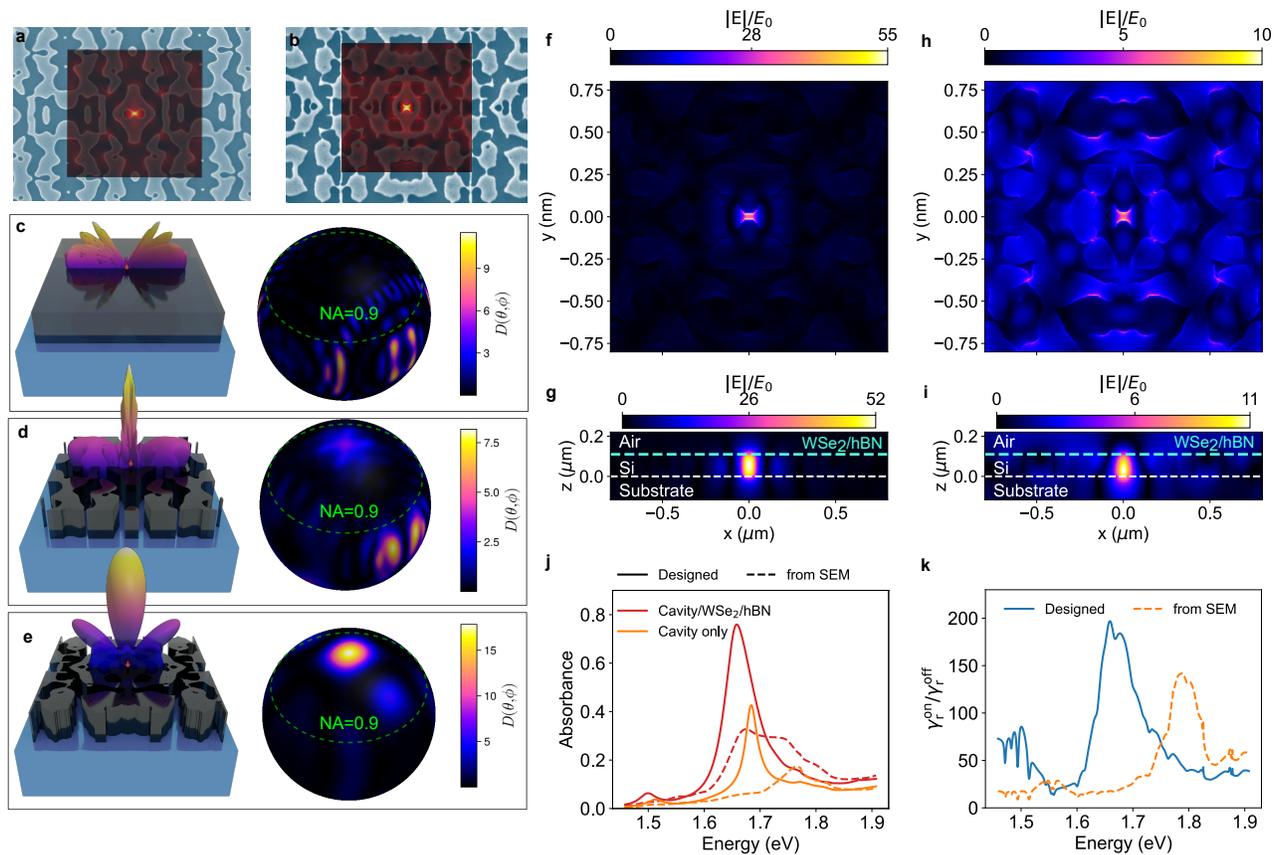

Figure 2. **Simulation results for topology-optimized dielectric cavities.**,**a,b**, The simulated near-field distributions overlaid on two example cavity designs, highlighting the locations of localized hotspots. **c**, Far-field radiation pattern (left) and radiation directivity plot (right) for a dipole placed off the cavity. The green dashed line indicates the collection numerical aperture (NA) of the microscope objective. **d,e**, Same as (**c**), but for the two topology-optimized cavities shown in (**a**) and (**b**), respectively. **f,g**, The simulated electric near-field distributions in the $xOy$ and $xOz$ planes for a unit cell of the cavity design in (**b**) as intended in the simulation. **h,i**, Same as (**f**) and (**g**), but using the pattern contours extracted from the experimental SEM images. These results incorporate fabrication distortions and illustrate their impact on near-field enhancement. **j**, the simulated absorbance spectra of the cavity design in (**b**), with and without the $WSe_2$/hBN stack, shown both for the ideal design (the solid lines) and the SEM-extracted structures (the dashed lines). The fabricated cavities exhibit a blue shift relative to the design and a red shift when loaded with the $WSe_2$/hBN stack. **k**, the simulated radiative emission rate enhancement relative to the off-cavity case, for the ideal and SEM-extracted structures.

To further characterize the origin of the non-exponential decay dynamics, we fitted the data using several models: triple-exponential, stretched-exponentials [93], a combined (mixed) model consisting of a stretched-exponential plus a single exponential, and the models describing the biexcitonic recombination. All fitting was performed after deconvolving each decay model with the instrument response function (IRF) to ensure physical accuracy [94] (for details of the fitting and deconvolution procedures, see III and IV). Among the tested models, the mixed model provides the best fit and yields consistent results for the power dependence of the decay parameters. This model is given by $I(t) = A_1 \exp(-(t/\tau_1)^\beta) + A_2 \exp(-t/\tau_2)$, where the second exponential term accounts for the long-time tail of the decay curve. The stretched exponential decay curves have been widely used [85, 86, 88, 93], following the original proposal by Förster [84], to describe the non-exponential dynamics in systems where the energy transfer occurs due to interactions. In this framework, the stretching parameter $\beta$ encodes the distinct physical characteristics of the system. It depends on the dimensionality of the system and the microscopic mechanism of the interaction. Specifically, $\beta$ is often expressed as $\beta = d/S$, where $d$ is the spatial dimensionality of exciton confinement and $S$ characterizes the type of interaction. For the dipole-dipole energy transfer, $S = 6$, which is expected to be the dominant mechanism for exciton interactions at room temperature. In our case, the excitons are confined to a two-dimensional WSe$_2$ monolayer, so $d = 2$, leading to $\beta = 1/3 \approx 0.333(3)$. Figure 3e shows the power dependence of $\tau_1^{\text{off}}/\tau_1^{\text{on}}$ while $\tau_2^{\text{off}}/\tau_2^{\text{on}}$ from Fig. 3f extracted during this fitting. As expected, we see around ×8 lifetime shortening on-cavity compared to the off-cavity case, which is consistent with the observation of the PL enhancements (Fig. 1). At relatively low powers, the stretching exponent for the off-cavity case is $\beta \approx 0.8$, and the decay dynamics are nearly exponential (Fig. 3g). As the pump power increases, however, $\beta$ decreases, resulting to progressively more non-exponential decay behavior. The observed decrease of $\beta$ with increasing excitation power suggests that the exciton-exciton interactions contribute to the non-exponential decay dynamics. At higher pump powers, the density of photoexcited carriers and excitons within the illuminated region increases, enhancing the likelihood of many-body processes such as exciton-exciton annihilation and interaction-mediated energy transfer. On the cavity, the $\beta$ values remain very small, $\beta \approx 0.4$, indicating the extreme non-exponential character of decay dynamics even at the lowest pump powers. The extracted value $\beta \approx 0.38$ is also very close to $\beta = 1/3 = 0.33$ for the expected dipole-dipole interactions. This observation further supports our earlier assumption that the exciton-exciton interactions play an important role in the decay dynamics. On the cavity, the optical near fields are tightly confined within nanoscale regions, leading to substantially stronger cavity-mediated resonant exciton-exciton interactions even at relatively low pump powers.

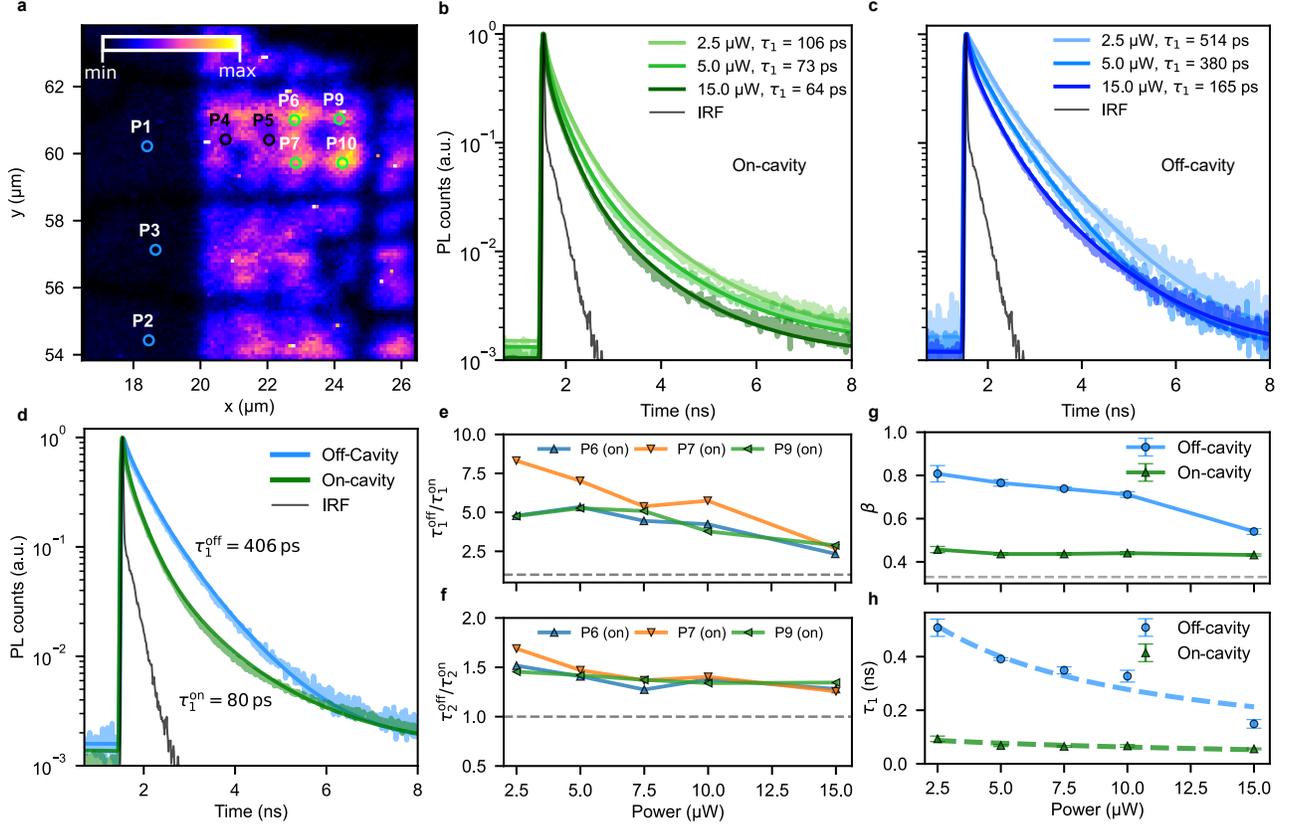

Figure 3. **Decay dynamics of** WSe$_2$ **excitons on and off the cavity arrays. a**, PL map showing selected measurement points on and off the cavity. **b**, the decay curves (green) from a cavity location (Point 10) at varying input powers, with corresponding mixed-model fits. The gray trace shows the instrument response function (IRF). **c**, the decay curves (blue) from a reference location off the cavity (Point 1), with corresponding fits and IRF as in (**b**). **d**, the averaged decay curves and fits on (blue) and off (green) the cavity, demonstrating the average ×5 lifetime reduction on the cavity. **e**, the ratio of short decay times $\tau_1^{\text{off}}/\tau_1^{\text{on}}$ for selected points as a function of pump power. **f**, Same as (**e**), but for the long decay time $\tau_2$ from the mixed-model fits. **g**, Stretching factor $\beta$ as a function of pump power on and off the cavity. **h**, the averaged short decay time $\tau_1$ on- and off- cavity as a function of the pump power, showing a nonlinear decrease as we pump the system stronger.

The extracted short decay time $\tau_1$ as a function of the pump power was fitted from these data using a power-dependent effective decay model (see SI for details), which incorporates different recombination channels through their density dependence(Fig. 3h). We find that the best agreement with the experimental data is achieved using a a the biexcitonic recombination model, described by

$$\tau_{\text{eff}} = \frac{\tau_{sp}}{1 + \tau_{sp} BG P_{\text{in}}},$$

where $G$ denotes the proportionality constant between the initial exciton density and the pump power ($n_0 = GP_{\text{in}}$ in the weak-excitation limit), $\tau_{sp} = 1/A$ is the spontaneous recombination lifetime ($A$ the spontaneous recombination rate), and $B$ is the biexcitonic recombination coefficient. We find that the biexcitonic recombination rate on the cavity is more than an order of magnitude larger than that of off-the-cavity case ($B_{\text{on}}/B_{\text{off}} \approx 33$). This result further confirms that the cavity enhances exciton–exciton interaction effects, and consequently biexcitonic recombination rates, by concentrating the optical near fields into nanoscale hotspots and thereby increasing the local exciton density in their vicinity.

Another clear indication of lifetime shortening results from the comparison of the low-temperature spectra of various excitonic species for the on and off-cavity array. Temperature-dependent PL spectra of hBN-encapsulated $WSe_2$ from T = 5 K to T = 200 K for the on and off cavity cases show multiple excitonic peaks at cryogenic temperatures (Fig. 4a,b), which is consistent with the previous reports in the literature [70, 77–83]. Notably, the excitonic resonances on the cavity are sharper and exhibit significantly narrower linewidth than those for the off the cavity case, reflecting the enhanced radiative decay in the confined cavity mode. This behavior is consistently observed across all the measured samples (see IV Fig. S5, Fig. S6, and Fig. S7). It is important to note that the observation of multiple excitonic species in the literature typically requires very clean, hBN-encapsulated samples under gate controlled doping. In contrast, our samples suffer from defects, and the some of the peaks become clearly resolvable only when the cavity-enhanced radiative processes narrow their linewidths.

Importantly, we detect distinct trion peaks in the low-temperature spectra. Trions have been previously reported in gate-controlled experiments [81] and occur only in $n$-doped $WSe_2$. Their presence here indicates unintentional doping, likely introduced during the Au-assisted transfer process [89]. The unintentional doping density can be estimated from the measured exciton-trion energy separation. In the measured spectra, we extract $E_{X_0} = 1.739$ eV and $E_{X_-} = 1.705$ eV (Fig. 4e), corresponding to an exciton-trion splitting of $\Delta E \approx 34$ meV.

Following the established relation between $\Delta E$ and the Fermi energy of the electron gas [81], the trion binding energy sets the characteristic scale for dissociation into a free exciton and an electron. From the low-temperature measurements (Fig. 4), we extract the dissociation temperature of $60 - 70$ K ($\sim 5 - 6$ meV ). Together with the measured exciton-trion splitting, this implies the Fermi energy on the order of several tens of meV, corresponding to an electron density of $n \sim 10^{11}$-$10^{12}$ cm$^{-2}$. In addition to trions, we also observe biexcitonic features, along with two additional low-energy peaks that are commonly attributed in the literature to defect-bound excitons [95]. In our measurements, these low-energy features persist even at room temperature, while trion and bi-exciton peaks vanish at elevated temperatures.

To investigate the behavior of these excitons, we performed power-dependent PL measurements at 5 K, both on and off the cavity (Fig. 4c and Fig. 4d, respectively). The spectra were analyzed using the Gaussian fits for the individual peaks, enabling us to quantitatively track their evolution with the excitation power (for fit comparisons, see IV Fig. S8). An example fit is presented in Fig. 4e for an input power of $P_{\text{in}} = 30 \, \mu$W. The excitonic resonances undergo negligible spectral shifts even at the excitation powers exceeding $300 \, \mu$W (Fig. 4c,d). At the same time, several peaks broaden markedly at higher powers and eventually merge with neighboring resonances. In contrast, the fitted peak amplitudes follow distinct power-law scaling behaviors (Fig. 4e): bright and brightened excitons scale nearly linearly ($\alpha \approx 1$), biexcitons scale quadratically ($\alpha \approx 2$), and biexciton complexes exhibit intermediate scaling with the exponents in the range $1 < \alpha < 2$.

Since the low-energy bound excitonic modes persist up to room temperature, we performed energy-filtered lifetime measurements to investigate their role in the decay dynamics. Representative decay curves were collected for the on (green) and off (blue) cavity cases (Fig. 4g) utilizing the tunable femtosecond pulsed laser that was set to near-resonant excitation at the main exciton peak (730 nm). The PL was spectrally filtered using 10 nm bandpass filters, which were tilted as needed to selectively transmit bandwidths centered at 750, 760, and 780 nm (see IV Fig. S9(e) for the measured filter responses at different angles). Clear differences in decay times are observed between the on- and off-cavity regions. At the same collection point, the variation of the detection wavelength - 750, 760, and 780 nm - produces noticeable changes in the extracted dynamics. For reference, the decay times were also obtained using a 550 nm long-pass (LP) filter that collects the full PL response. These trends are summarized in Fig. 4h, where the fitted short decay components are plotted as a function of the detection wavelength. Interestingly, while the bound exciton emission centered at 780 nm would normally be expected to decay slower than the bright exciton emission at 750 nm, our data reveal an unexpected $\sim 2$ times shorter decay time for the bound exciton compared to the bright exctions. As expected, the average decay times are also ∼ 2 times shorter in the cavity than in the off-cavity for the exciton emission at all wavelengths.

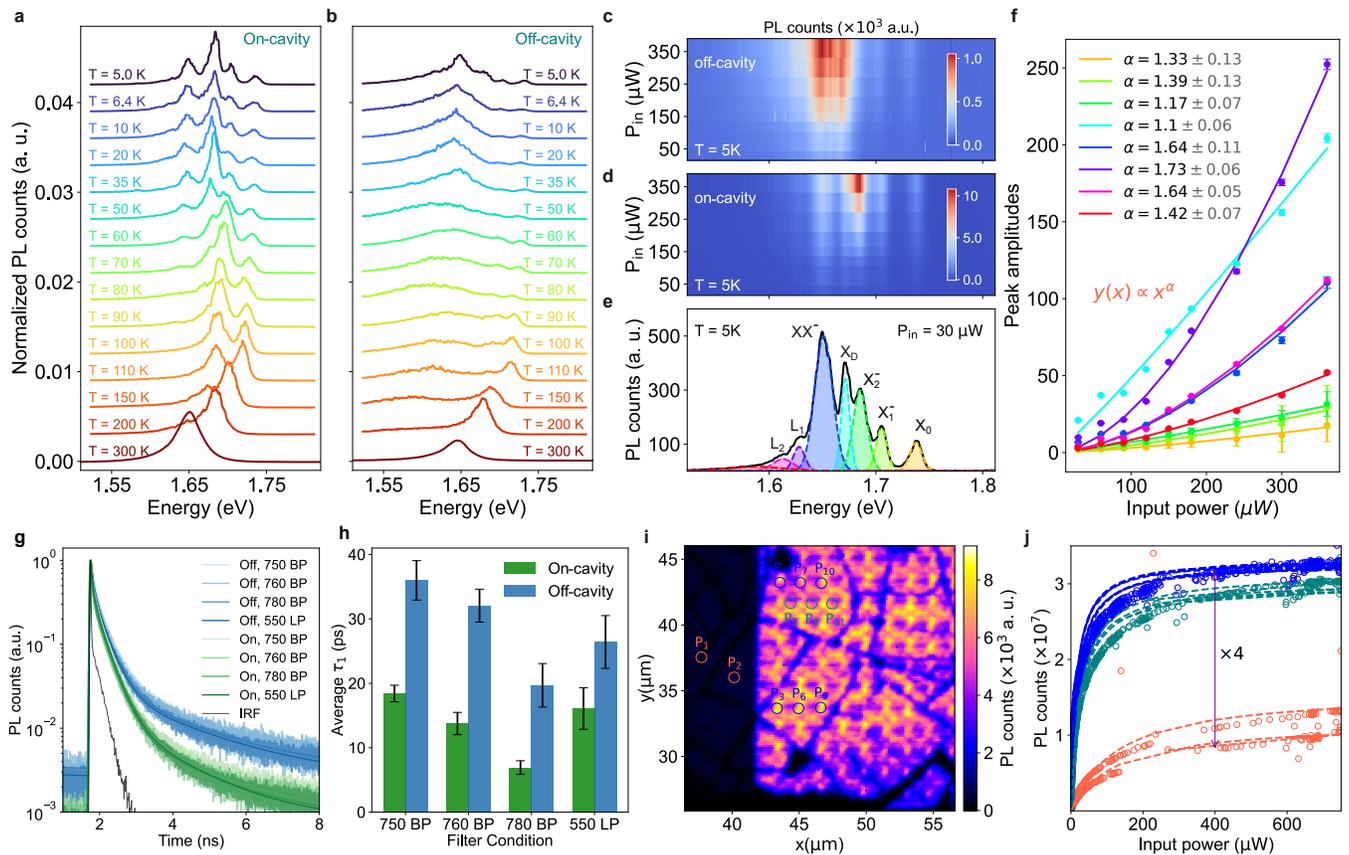

Figure 4. **Low temperature spectra, power and temperature dependence fits.** **a**, on and **b**, off cavity temperature dependent WSe$_2$ spectra at temperatures from T = 5.0 K to T = 200 K. **c**,**d**, the power dependent density plots of WSe$_2$ spectra for the off and on the cavity cases, respectively, collected at 5 K fixed temperature. At elevated excitation powers, some excitonic resonances exhibit pronounced broadening, leading to their merging with adjacent peaks. **e**, the example spectra at 5 K and for input power P$_{in}$ = 30 $\mu$W and their Gaussian fitting. The peak assignment is based on the comparison of the measured peak energies and their relative shifts with values reported in the literature. **f**, the scatter data with the error bars are the peak amplitudes extracted from the fits as a function of the input power. The error bars represents the fit errors. The solid lines show the power low $A_j \propto P_{in}^{\alpha}$ fits to these data. The power $\alpha$ and respected fit errors are shown as a legend for each individual peak. **g**, the energy filtered lifetime measurements at room temperature on and off the cavity. **h**, the fitted fast decay times $\tau_1$ for the on and off cavity cases as extracted from the fitting of the data in (**g**). **i**, the reference PL map to show the specific points for which the saturation curves in (**j**) are measured. **j**, the measured saturation curves on (the blue and green scatter data points) and off the cavity (the orange points).

Finally, we measured the saturation behavior of the excitonic emission on and off the cavity. A clear difference is observed (Fig. 4i,j). On the cavity, the saturation curve rises more steeply and reaches the saturation at ∼ 20$\mu$W, whereas off-cavity the response is slower and saturates at higher powers (∼ 100$\mu$W). The faster saturation observed on the cavity is consistent with the enhanced exciton-exciton interactions, further supporting that the field localization strengthens the interaction effects and can yield substantially larger excitonic nonlinearities. A detailed study of the resulting enhancement of two-photon nonlinearities is left for a follow-up work.

## III. METHODS

### A. Adjoint Topology Optimization of Metasurface Cavities

We employ a gradient-based topology-optimization method [37, 49, 51] in tandem with three-dimensional finite-difference time-domain (FDTD) simulations to obtain silicon-on-sapphire cavity unit cells that confine optical modes deep below the diffraction limit. The design region is a single metasurface unit cell with periodicity $a \approx 2\lambda_0$ and $\lambda_0/(2n_{Si})$ thickness, which in our experiments ranges between 100-200 nm. The unit cell is bounded by PEC/PMC boundary conditions, which allows one to leverage the symmetric nature of the design and simulate only a quarter of the simulation do-main. The design region is filled with a continuous material-density field $\rho(x,y) \in [0,1]$, which is interpolated into the dielectric permittivity through

$$\varepsilon(x,y) = \varepsilon_{\text{air}}[1 - \rho(x,y)] + \varepsilon_{\text{Si}}\rho(x,y)$$

where $\rho = 0$ corresponds to air and $\rho = 1$ to crystalline silicon. The figure of merit guiding the search is the electric-field intensity at the central surface point

$$F = \left| E(0,0,z_{\text{surface}}) \right|^2,$$

whose topological gradient is obtained via the adjoint method. Writing objective function as $F = \int_\chi |E|^2 \, d^3x$ over a small evaluation region $\chi$, a first-order perturbation of the permittivity gives

$$\delta F = 2 \operatorname{Re} \int_\chi E^*(x) \, \delta E(x) \, d^3x.$$

Expressing $\delta E$ with the electric Green tensor and invoking reciprocity yields

$$\delta F = 2 \operatorname{Re} \int_\psi \mathbf{P}^{\text{ind}}(x') \cdot \mathbf{E}^A(x') \, d^3x',$$

where $\mathbf{P}^{\text{ind}}(x') = \Delta\varepsilon(x') \, \mathbf{E}^{\text{fwd}}(x')$ is the polarization density from the forward field and the adjoint field $\mathbf{E}^A$ is produced by a point electric dipole of complex amplitude $\partial F/\partial \mathbf{E} = 2 \mathbf{E}^*$ placed at the target location. Thus, each optimization step comprises a forward FDTD run to obtain $\mathbf{E}^{\text{fwd}}$, an adjoint run driven by the dipole source to evaluate $\mathbf{E}^A$. Gradient scaling and update are performed using the adaptive moment estimation method (ADAM)[96]. The topology-optimization loop updates the continuous density field with adaptive moment gradient steps computed from adjoint sensitivities. At each iteration, the design variable $\rho$ is filtered by convolution with a Gaussian kernel in Fourier space with 20 nm radius to suppress features below the minimum fabricable size. The filtered design field then undergoes a smoothed Heaviside projection whose sharpness parameter $\beta$ is gradually ramped from 1 to 1000, systematically pushing $\rho$ toward binary values and enforcing manufacturability. We iterate this sequence-forward solve, adjoint solve, filtered ADAM update, and binarization for roughly 200-250 iterations, based on FOM saturation and design field binarization.

### B. Fabrication

A 100 nm-thick epitaxial silicon-on-sapphire (SoS) wafer is diced into 6.5 by 6.5 mm pieces. The pieces are then cleaned by ultrasonication in acetone, isopropanol, and deionized water, and dried with dry $N_2$. The sample surface is dehydrated by heating in a nitrogen atmosphere and primed with hexamethyldisilazane (HMDS) vapor in a Yield Engineering Services YES LP-III vacuum oven.

The sample is spin-coated with a 50 nm thick film of EBL resist (ZEP 520A diluted 1:2 in anisole) at 4000 rpm for 45 s, and the coated sample is baked on a hotplate at 180°C for 3 min. 100 by 100 $\mu m^2$ meta-cavity arrays are written in the electron beam resist using an Elionix ELS-BODEN 150 EBL system with a 150 keV electron beam at a current of 200 pA and base exposure dose of 325 $\mu C/cm^2$. Multiple copies of the array are written with various combinations of unit cell scaling (detuning) and proximity effect correction (PEC) parameters. The sample is developed in chilled (approx. 3°C) o-xylene for 30 s, rinsed in isopropanol for 30 s, and dried with dry $N_2$.

The developed pattern is transferred to the underlying silicon by inductively coupled reactive ion etching (ICP-RIE) in an SPTS Omega LPX Rapier reactive ion etcher. After etching, the remaining electron beam resist is removed by rinsing in acetone and isopropanol, followed by 4 min of $O_2$ plasma etching in a Matrix Integrated Systems 105 plasma asher.

### Flake transfer

Monolayer $WSe_2$ flakes were prepared using a gold-template-assisted mechanical exfoliation technique adapted from Wu *et al.* [89]. Bulk $WSe_2$ crystals were exfoliated onto a patterned Au grid on a $Si/SiO_2$ substrate immediately after broad Ar ion beam polishing. Broad ion beam polishing was performed using a JEOL IB-19500CP Cross Section Polisher with the beam parallel to the substrate surface. The procedure recovers the Au surface and enables enhanced exfoliation yield and uniformity similar to those of freshly deposited Au thin films. Monolayer regions were identified via optical contrast and subsequently transferred along with an h-BN capping thin flake onto our patterned Si substrate using a dry transfer method with a polycarbonate (PC)/polydimethylsiloxane (PDMS) stamp. After the transfer, the samples were soaked in chloroform to remove residual PC and improve adhesion. Monolayer thickness was confirmed by photoluminescence spectroscopy and Raman spectroscopy. Original samples in the experiment that had flake thickness of a few to tens of nanometers were mechanically exfoliated from bulk crystals using a standard adhesive tape method and their thickness was measured with atomic force microscopy (AFM). These thicker flakes were also transferred onto the patterned Si substrate using a dry transfer method with a polycarbonate (PC)/PDMS stamp.

### C. Optical measurements

All optical measurements were performed using custom-built scanning confocal microscopy setups. Room-temperature characterization was conducted on an inverted microscope platform, while low-temperature measurements used a top-access closed-loop helium cryostat equipped with a vacuum-compatible, high numerical aperture (NA) objective. For the room temperature setup, sample scanning was achieved using a nanopositioning stage. For the low-temperature setup, a 4F scanning technique using dual-axis scanning galvo system was utilized. The excitation and PL collection for both setups were configured in a reflection geometry. A detailed schematic and component list for each setup can be found in Fig. S1 and Fig. S10 and IV S1.

### Continuous-Wave PL, Spectral Analysis, and Saturation Measurements

For continuous-wave (CW) experiments, samples were excited with a 532 nm diode laser. The collected PL was passed through a set of long-pass and band-pass filters to isolate the emission band centered at 750 nm. The signal was then spatially filtered with a confocal pinhole and detected by a single-photon avalanche diode (SPAD). Saturation curves were obtained by measuring the steady-state PL intensity as a function of excitation power. For spectral analysis, the PL was directed to a visible-to-near-infrared spectrometer, and all spectra were corrected for detector response and background.

### Time-Resolved PL Measurements

Time-resolved PL (TRPL) measurements were performed using a tunable femtosecond Ti:Sapphire laser for both above-gap (520 nm, via second harmonic generation) and near-resonant (730 nm) excitation. The fluorescence decay traces were recorded using a high-resolution SPAD and a time-correlated single-photon counting (TCSPC) module.

### Data Acquisition and Analysis

All measurements were controlled and synchronized using custom LabVIEW software. The resulting fluorescence decay curves were analyzed using nonlinear least-squares fitting routines in Julia and Python. To accurately model the observed dynamics, the data were fit to several models, including single exponential, biexponential, and stretched exponential forms, after deconvolution with the instrument response function (IRF). The full mathematical descriptions of the models and the fitting procedure are detailed in IV S4.

## IV. DISCUSSION AND CONCLUSION

We showed that topology-optimized dielectric cavities provide an effective platform for controlling light–matter interactions in two-dimensional semiconductors. By harnessing adjoint-based inverse design, we realized CMOS-compatible silicon-on-sapphire cavities with deeply subwavelength mode volumes and engineered far-field directivity, enabling deterministic coupling to excitons in monolayer $WSe_2$. Photoluminescence mapping revealed the enhancements approaching an order of magnitude relative to off-cavity, in quantitative agreement with simulations when fabrication tolerances are included. Importantly, the cavity coupling enables not only the amplified emission but also reshapes the excitonic dynamics. Specifically, the low-temperature spectroscopy showed sharpened excitonic resonances, including bright excitons, trions, and biexcitons, while the time-resolved measurements revealed cavity-mediated lifetime shortening and non-exponential decay signatures, which are indicative of the exciton-exciton interactions. These results provide direct evidence that the developed cavities accelerate radiative processes and mediate correlated exciton phenomena.

In the broader context of the reported PL enhancements in 2D materials and related platforms (Table I), the observed results are very competitive and the enhancement values are comparable to those in previously reported plasmonic structures. Although reported enhancement values in the literature span several orders of magnitude, much of this variation arises from the specific normalization conventions used in different experiments. For example, in plasmonic systems the enhancement is often defined relative to the ratio between the excitation beam area and the nanoscale hotspot where most of the field localization occurs. In the tip-enhanced spectroscopy, geometric effects such as the tip radius-to-distance ratio play a dominant role. To enable more systematic comparison, in Table I we list both the claimed enhancement factors as reported in each study and the raw enhancement values that can be directly extracted from PL intensity or spectra peak amplitudes, while noting the cases where the geometric corrections could not be reconstructed. As specific examples, in Ref.[13] the enhancement factors were extracted from integrated PL counts without background subtraction; a more rigorous analysis based on the peak fitting and the amplitude comparison would likely yield substantially smaller values. Similarly, the exceptionally large enhancement reported in Ref. [16] appears to originate from broadband cavity effects that simultaneously enhance both pump absorption and emission. This leads to a multiplicative contribution nearly an order of magnitude larger than other studies, including our own, where the enhancement is primarily governed by the Purcell effect in the emission channel. We also note that several recent preprints have reported dielectric bound-state-in-the-continuum (BIC) cavities for PL enhancement. However, the quantitative claims in these works are difficult to evaluate, as the reported figures and plots lack vertical axis values for the raw PL signals. Due to this absence of verifiable quantitative data, we have not included these studies in the present comparison.

Table I. PL and emission rate enhancement

| Year | Reference | Claimed Enhancement | Raw Enhancement | Geometry Factor |
|------|-----------|---------------------|-----------------|-----------------|
| 2025 | this work | $10^3$ | 10 - 14 | 50-100 |
| 2025 | Ref [32] | 5 | 5 | - |
| 2025 | Ref [6] | $2.1 \cdot 10^4$ | 14.23 | 1479.45 |
| 2024 | Ref [7] | $1.8 \cdot 10^4$ | 7.44 | 2419.75 |
| 2023 | Ref [8] | $1.6 \cdot 10^3$ | 1.1 | 1444 |
| 2023 | Ref [9] | $9 \cdot 10^4$ (AS), $4 \cdot 10^3$ ($X_D$), $9 \cdot 10^2$ ($X_0$) | 20 (AS), 12 ($X_D$), 3 ($X_0$) | 50-430 |
| 2023 | Ref [10] | $6 \cdot 10^3$ ($X_{IL}$) | - | tip-enhanced |
| 2022 | Ref [11] | - | 1.5 | - |
| 2018 | Ref [12] | $1.7 \cdot 10^3$ | 7.5 | 225 |
| 2018 | Ref [13] | $0.7 \cdot 10^3$ | 700 | - |
| 2018 | Ref [36] | 26 | 26 | - |
| 2018 | Ref [21] | 7 | 7 | - |
| 2018 | Ref [14] | 13(PL), 57(LT) | 13(PL), 57(LT) | - |
| 2017 | Ref [15] | 200 | 20 | 10 |
| 2016 | Ref [16] | $2.0 \cdot 10^4$ | 1800 | 11.31 |
| 2015 | Ref [28] | 37 (LT) | - | - |

[a] AS = Anti-Stokes; $X_D$ = Dark exciton; $X_0$ = Bright exciton; $X_{IL}$ = Interlayer exciton; PL = Photoluminescence enhancement; LT = Lifetime shortening (Purcell factor).

Overall, our findings establish topology-optimized dielectric cavities as a versatile, CMOS-compatible platform to plasmonics for tailoring exciton-photon and exciton-exciton interactions. Unlike plasmonic architectures, these dielectric cavities achieve deep subwavelength confinement without ohmic loss, thereby preserving coherence while enabling efficient out-coupling. Beyond the enhanced PL, the combination of the strong near-field localization and controlled radiation offers a path toward unprecedented regimes of exciton nonlinearities, valley-selective emission, and cavity-enabled quantum many-body physics. Nevertheless, scalable fabrication of these structures presents significant fabrication challenges, including the proximity effects in EBL, shape distortions, and roughness due to reactive ion etching (RIE), which must be carefully addressed to preserve the designed optical performance. We anticipate that this platform will accelerate the integration of van der Waals materials into scalable quantum photonics, bridging the gap between fundamental exciton science and on-chip quantum technologies.

## SUPPLEMENTARY INFORMATION

The Supplementary Information is included as an appendix at the end of this submission.

## DATA AVAILABILITY

The data that support the findings of this study have been included in the main text and Supplementary Information. All other relevant data supporting the findings of this study are available from the corresponding authors upon request.

## CODE AVAILABILITY

The codes used for the lifetime and spectra fitting are available from the corresponding authors upon reasonable request.

## ACKNOWLEDGMENTS

This work was supported by the Office of Naval Research (ONR) (Grant No. 13001182). This work was performed in part at the Harvard University Center for Nanoscale Systems (CNS); a member of the National Nanotechnology Coordinated Infrastructure Network (NNCI), which is supported by the National Science Foundation under NSF award no. ECCS-2025158.

## AUTHOR CONTRIBUTIONS

O.M. developed the fabrication methods and fabricated most metacavities. B.T. performed room-temperature measurements, developed software for experiment control and automation, analyzed the data, prepared samples, and contributed to manuscript writing. O.Y. developed the numerical methods and designed the metacavities. D.C. contributed to fabrication. K.P. and M.Y. assisted with data analysis and manuscript preparation, with M.Y. also contributing to low-temperature measurements. A.E.L.A. assisted with monolayer flake transfer. H.A. and A.B.S. designed and constructed the low-temperature setup. Y.P.C. provided access to facilities. H.A. provided theoretical guidance; A.V.K. supervised topology optimization work of O.Y.. V.M.S., F.C., and A.B. supervised the project, provided resources, and participated in discussions. V.M. conceived the project, supervised all aspects, performed simulations, assisted with experiments and sample preparation, and coordinated the manuscript writing.

V.M., B.T., K.P., and M.Y. wrote the manuscript with input from all the authors.

### Corresponding authors

Correspondence to Vahagn Mkhitaryan.

### ETHICS DECLARATIONS

### Competing interests

The authors declare no competing interests.

# Topology-Optimized Dielectric Cavities for Enhanced Excitonic Light Emission from WSe₂

Owen Matthiessen,[1,2,*] Brandon Triplett,[1,2,*] Omer Yesilyurt,[1,2,*] Davide Cassara,[3,*] Karthik Pagadala,[1,2] Morris M. Yang,[1,2] Andres E. Llacsahuanga Allcca,[4] Hamza Ather,[4,1,2] Colton Fruhling,[1,2] Abhishek Bharatbhai Solanki,[1,2] Yong P. Chen,[1,5,2,4] Hadiseh Alaeian,[1,5,2,4] Alexander V. Kildishev,[1,2] Vladimir M. Shalaev,[1,5,2,4] Federico Capasso,[3] Alexandra Boltasseva,[1,5,2,†] and Vahagn Mkhitaryan[1,2,‡]

[1]*Elmore Family School of Electrical and Computer Engineering, Purdue University, West Lafayette, IN 47906, USA*
[2]*Birck Nanotechnology Center, Purdue University, West Lafayette, IN 47906, USA*
[3]*John A. Paulson School of Engineering and Applied Sciences, Harvard University, Cambridge, MA 02138, USA*
[4]*Department of Physics and Astronomy, Purdue University, West Lafayette, IN 47907, USA*
[5]*Purdue Quantum Science and Engineering Institute, Purdue University, West Lafayette, IN 47907, USA*

## Contents

---

*These authors contributed equally.

†Corresponding author: aeb@purdue.edu

‡Corresponding author: vmkhitar@purdue.edu

# S1. LOW TEMPERATURE OPTICAL SETUP

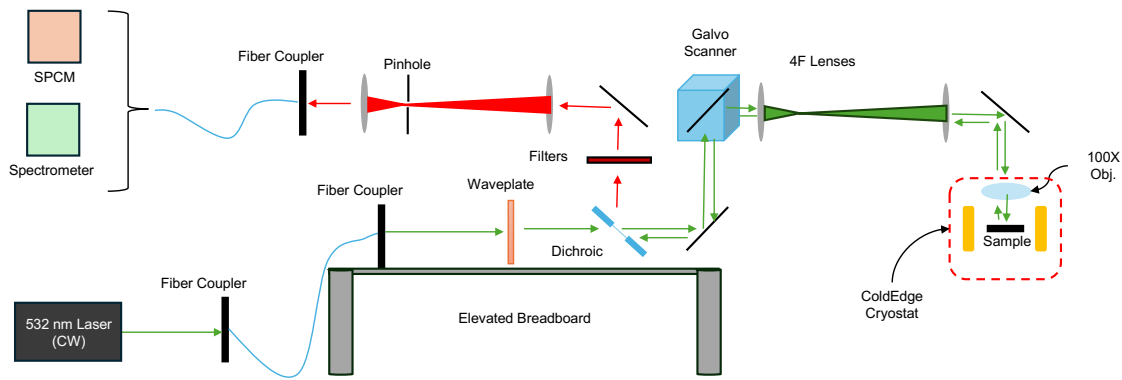

FIG. S1: **Schematic of low-temperature optical setup.** A 532 nm CW laser (Cobolt) is focused onto the sample inside a closed-loop helium cryostat via a high-NA objective (Attocube). Scanning is performed with a 2D galvo system. The collected PL is spatially filtered with a 50 µm pinhole and sent to one of two detection arms: a SPAD (Excelitas) for photon counting or a spectrometer (Princeton Instruments) with a liquid-nitrogen-cooled CCD for acquiring emission spectra.

Low-temperature PL measurements were performed using a custom-built cryogenic confocal microscope (Fig. S1). The setup features a ColdEdge Stinger Hydra cryostat with a closed-loop liquid helium cooling system. Excitation and collection were achieved with a low-temperature compatible objective (Attocube, LT-APO/532-RAMAN/0.82 NA) mounted inside the cryostat. Sample scanning was performed using a 2D large-beam galvo system (Thorlabs, GVS012).

The excitation source was a fiber-coupled 532 nm CW laser (Cobolt 08-DPL, Hubner Photonics). Collected PL was spatially filtered with a 50 µm pinhole before detection. The signal was directed either to a SPAD (Excelitas, SPCM-AQRH) for photon counting or to a Princeton Instruments HRS-300 spectrometer for spectral characterization. The spectrometer was equipped with interchangeable diffraction gratings (1200 g/mm and 600 g/mm) and a liquid-nitrogen-cooled PyLoN CCD camera for low-noise spectral acquisition.

## S2. RAMAN SPECTROSCOPY

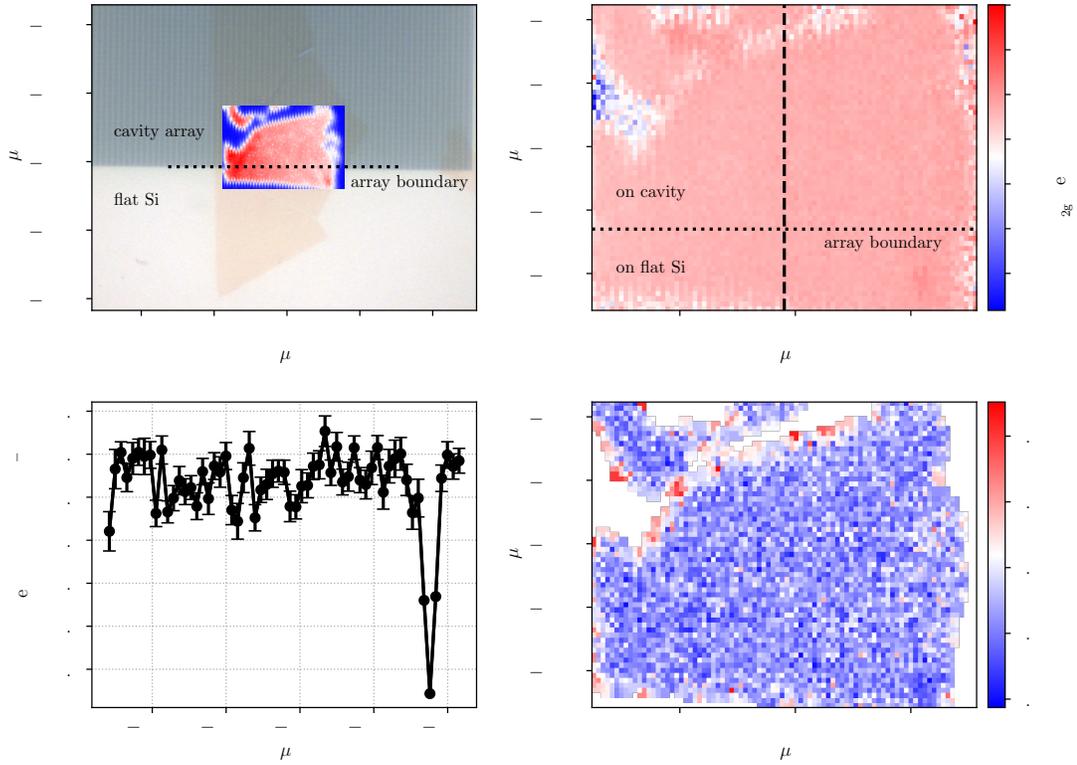

FIG. S2: **Raman mapping of a WSe$_2$ flake. a,b**, Map of the Raman E$^1_{2g}$ peak position superimposed on an optical microscope image. **c**, A line profile of the peak position across the flake, indicating strain variation. **d**, Fit errors from the analysis.

Raman maps were acquired using a Thermo Scientific DXR3xi Raman imaging microscope with a 532 nm excitation laser and a spatial resolution of 0.4 $\mu$m per pixel. Point spectra from the maps were fit using a dual Lorentzian oscillator model to extract the peak positions of the E$^1_{2g}$ and A$_{2g}$ modes of WSe$_2$. The peak positions were extracted from the resulting fits and the map of the E$^1_{2g}$ peak position superimposed on an optical microscope image is shown in Figure S2a. S2b shows the Raman E$^1_{2g}$ Peak position map, highlighting the areas where the WSe$_2$ flake is on-cavity and off-cavity. A dashed-line profile of peak positions is displayed in S3b and plotted in S2c. The fit errors shown in S2d clearly indicate the WSe$_2$ flake compared to the background.

## S3. ATOMIC FORCE MICROSCOPY

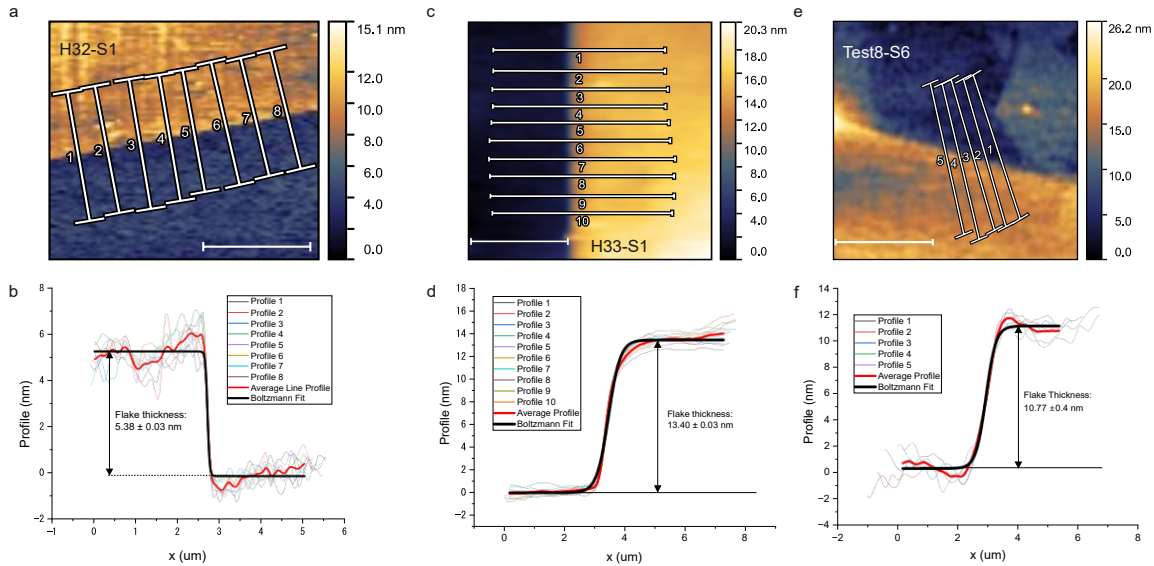

FIG. S3: **hBN thicknesses in the different samples as estimated using atomic force microscopy a,c,e**, AFM height maps of samples labeled H32-S1, H33-S1, and Test8-S6, respectively. White lines indicate the locations of individual height profiles used to determine the flake thickness. Scale bars are 5 μm. **b,d,f**, Height profiles along the lines shown in (**a,c,e**), with individual profiles, the average profile in red, and a Boltzmann fit to the average profile in black. The extracted flake thicknesses from the Boltzmann fits are indicated in each plot.

AFM scans were taken on 3 samples using Park Systems AFM under NCFM modes. Area scans were saved and processed on Gwyddion software for surface leveling and line scan fitting.

## S4. FABRICATION RESULTS

The small, dense features, and high fidelity demanded by the inverse design process place unique constraints on the fabrication process. The etching procedure, in particular, had to be carefully optimized. Small features require a thin etch mask, which in turn requires high etch selectivity; narrow gaps require straight sidewalls; and fabrication error sensitivity requires smooth sidewalls. We opted for a non-switched Bosch-type etch using $SF_6$, $C_4F_8$, and $CHF_3$ for simultaneous etching and sidewall passivation. The operating pressure, gas flows, ICP power, and bias power were optimized in order to balance sidewall passivation and horizontal facet fluoropolymer removal, yielding an etch with a selectivity of approximately 5:1 against the etch mask, and smooth sidewalls at an angle of 88 degrees. A meta-cavity array produced using the final optimized etch can be seen in S4a. Etching results produced by intermediate optimization steps and their respective etching parameters are shown in S4b-c.

A key technique used as part of the previously described optimization was automated contrast-based segmentation and contour detection. Secondary electron imaging is generally used to capture topography on the scale of the features in our meta-cavities. As a result, sharp corners, such as those at the edges of etched regions, exhibit a so-called "secondary electron glow." We developed a script utilizing thresholding and denoising algorithms to isolate the secondary electron glow present in our SEM images, thereby extracting the contours of the as-fabricated meta-cavities. Doing so allows for a direct comparison against the blueprint design, decreasing the optimization cycle time, providing valuable insights into the impact of processing variations on the final fabricated structure, and enabling the simulation of the as-fabricated structure. Examples of SEM images and the detected contours along with comparisons of the detected contours to the blueprint designs are shown in S4d-g.

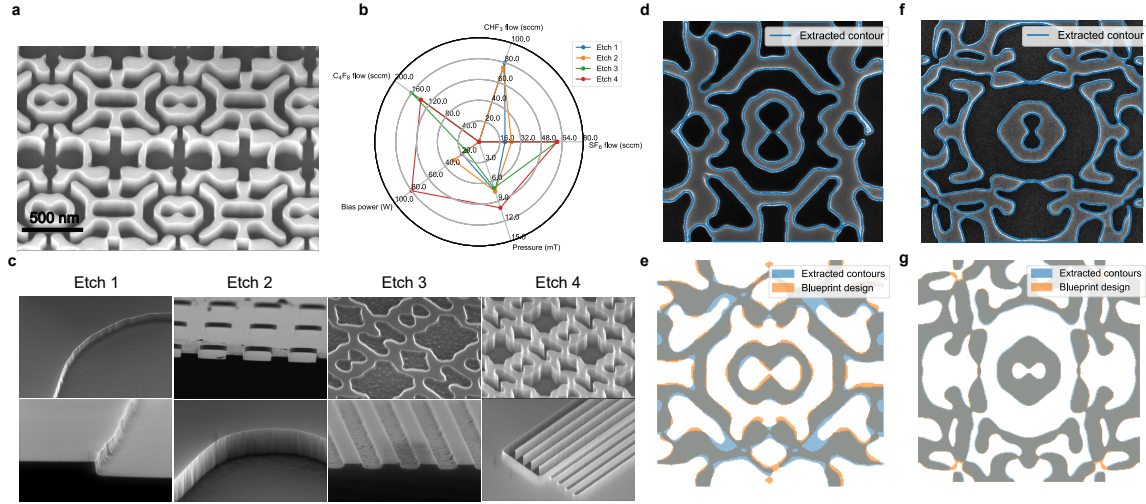

FIG. S4: **Etch Optimization and cavity shape distortions compared to original design**. **a**, representative SEM image of etched cavity. **b**, parameters of four selected etches from the optimization process. **c**, representative SEM images of four selected etches from the optimization process. **d,f**, SEM images of fabricated structure and detected contours. **e,g**, comparison of fabricated structure and blueprint design.

## S5.  LIFETIME

### S6.1 Measurement

The lifetime measurements were conducted using the room-temperature confocal microscope equipped with the femtosecond Ti:Sapphire laser and time-correlated single-photon counting (TCSPC) module as shown in Fig.  S5. The laser was operated at 1040 nm and frequency doubled to 520 nm. The excitation powers ranged from 1-50 $\mu$W. The collection time was set to 10 seconds. The instrument response function (IRF) of the system was measured separately by detecting scattered laser light from a metal post. Energy-filtered measurements were performed using ±10 nm bandpass filters centered at 750 and 760 nm. To access 780 nm, an 800 nm bandpass filter was tilted by 25°. The transmission spectra of the filters are shown in Fig.  S9e.  For measurements with the tilted 800 nm filter, the collection time was doubled from 10 to 20 s to compensate for reduced efficiency.

### S6.2 Fitting

The photoluminescence lifetime data were fit to several candidate models to identify the most accurate physical description of the decay process. The fitting was performed in Python using the `lmfit` package. We used a convolution-based fitting technique where each theoretical model was convolved with the measured instrument response function (IRF) before being compared to the experimental data. This approach is numerically more stable than direct deconvolution, especially for noisy signals.

The model parameters, such as the fast and slow lifetimes ($\tau_1$, $\tau_2$) and the stretching factor ($\beta$), were optimized using a weighted least-squares minimization routine. The fits were weighted by the Poisson noise of the photon counts ($\frac{1}{\sqrt{\text{counts}}}$) and used a strict convergence tolerance of $10^{-8}$. We tested three candidate models: (1) a multi-exponential, (2) a stretched exponential, and (3) a mixed model. The mixed model was ultimately selected for presentation in the main paper because it consistently yielded the best fits, as determined by reduced chi-squared values close to unity ($\chi^2_{\text{red}} \approx 1$) and a visual inspection of the residuals. Fits with poor statistics or systematic residuals were rejected from the final analysis. The model functions are listed below.

**Stretched exponential model:**

$$I(t) = A \exp\left[-\left(\frac{t}{\tau}\right)^{\beta}\right]$$

**Multi-exponential model (n=3):**

$$I(t) = A_1 e^{-t/\tau_1} + A_2 e^{-t/\tau_2} + A_3 e^{-t/\tau_3}$$

**Mixed model:**

$$I(t) = A_1 \exp\left[-\left(\frac{t}{\tau_1}\right)^\beta\right] + A_2 e^{-t/\tau_2}$$

**IRF model:**

$$I(t) = (y_x * y_g)(t) + y_0, \tag{S1}$$

$$y_x = A\left[1 + \tanh\left(\frac{t-\mu}{t_r}\right)\right]\left(e^{-\frac{t-\mu}{t_d}} + Be^{-t/\tau}\right), \tag{S2}$$

$$y_g(t) = \frac{1}{\sigma\sqrt{2\pi}}\exp\left[-\frac{(t-\mu)^2}{2\sigma^2}\right]. \tag{S3}$$

### S6.3 Supplemental data

The analogous figures for our other extensively measured samples (H33-S1 and Test-8 S6), corresponding to Figure 4 in the main text (H32-S1), are shown below.

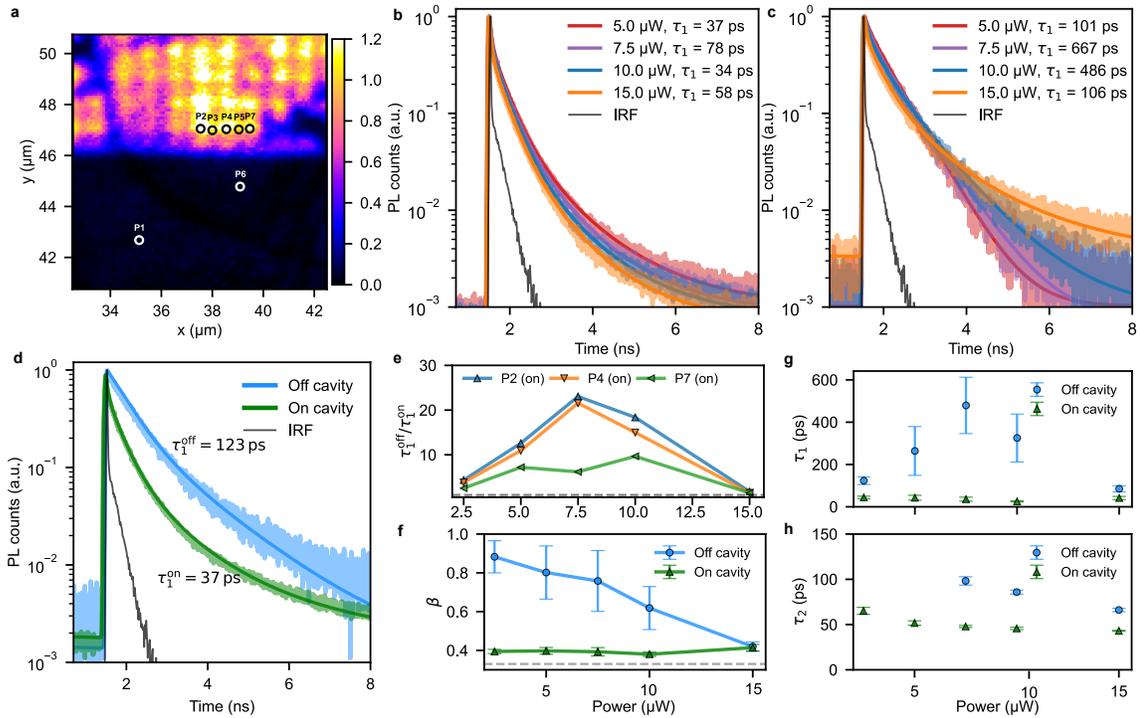

FIG. S5: **Lifetime fitting of non-exponential decay with mixed modeling for sample Test-8 S6. a**, PL map showing selected measurement points on and off the cavity. **b**, Decay curves from a cavity location (Point 2) at varying input powers, with corresponding mixed-model fits. The gray trace shows the instrument response function (IRF). **c**, Decay curves from a reference location off the cavity (Point 1), with corresponding fits and IRF as in (**b**). **d**, Averaged decay curves and fits on and off the cavity. **e**, Ratio of short decay times $\tau_1^{\text{off}}/\tau_1^{\text{on}}$ for selected points as a function of pump power. **f**, Stretching factor $\beta$ as a function of pump power on and off the cavity. **g**, Averaged short decay time $\tau_1$ on and off the cavity as a function of pump power, showing a nonlinear decrease off-cavity and a nearly linear dependence on-cavity. **h**, Same as (**g**), but for the long decay time $\tau_2$.

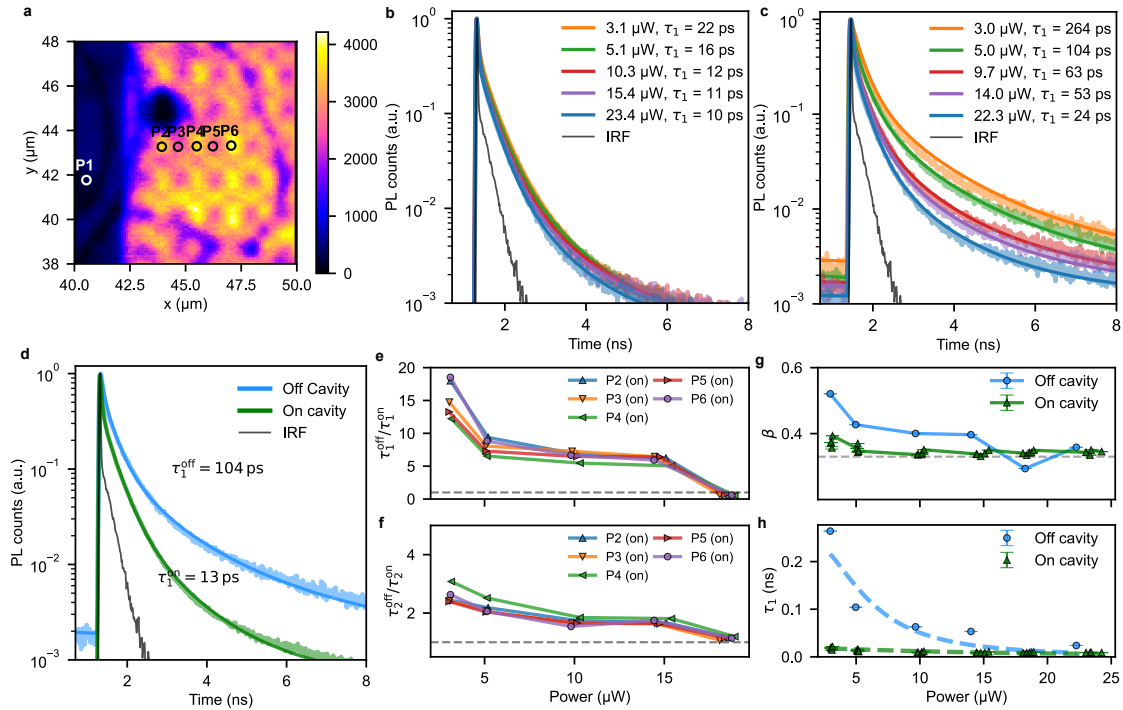

FIG. S6: **Lifetime fitting of non-exponential decay with mixed modeling for sample H33-S1. a**, PL map showing selected measurement points on and off the cavity. **b**, Decay curves from a cavity location (Point 6) at varying input powers, with corresponding mixed-model fits. The gray trace shows the instrument response function (IRF). **c**, Decay curves from a reference location off the cavity (Point 1), with corresponding fits and IRF as in (**b**). **d**, Averaged decay curves and fits on and off the cavity. **e**, Ratio of short decay times $\tau_1^{\text{off}}/\tau_1^{\text{on}}$ for selected points as a function of pump power. **f**, Same as (**e**), but for the long decay time $\tau_2$ from the mixed-model fits. **g**, Stretching factor $\beta$ as a function of pump power on and off the cavity. **h**, Averaged short decay time $\tau_1$ on and off the cavity as a function of pump power.

While multiple devices were characterized, the data from sample H32-S1 is highlighted in the main text. This sample provided the most comprehensive dataset for the off-cavity regions, allowing for the most reliable comparison between cavity-coupled and uncoupled emission dynamics.

We include the analogous data for two other representative samples (H33-S1 and Test 8 S6) in Supplementary Figs. S5 and S6. Although the off-cavity measurements for these samples contained fewer data points or showed larger fit uncertainty due to lower signal, the on-cavity decay dynamics were in excellent agreement across all measured devices, confirming the generality of our conclusions.

## S6.4 Spatially Resolved Lifetime Mapping

We performed spatially resolved lifetime measurements to map the variation of decay parameters across the sample surface. These maps were acquired by scanning the sample in 1-$\mu$m steps and recording a full photoluminescence decay curve at each pixel. Each decay curve was then fit using the mixed model described previously. The extracted fitting parameters (e.g., $\tau_1$, $\tau_2$, $\beta$) were then plotted as a function of position, as shown in Supplementary Fig. S9a,b, and c. The spatial resolution of these maps is likely limited by mechanical drift over the prolonged (multi-hour) acquisition period.

## S6.5 Model Validation via Reduced Chi-Squared Analysis

To quantitatively validate our choice of fitting model, we compared the reduced chi-squared ($\chi^2_{red}$) values for all three candidate models across our datasets, as shown in Supplementary Fig. S7. A fit was considered reliable and included in our statistical analysis only if it met the criterion of $\chi^2_{red} < 2$.

Fits that failed this criterion were predominantly from low-power measurements with a poor signal-to-noise ratio and were consequently excluded. The results clearly demonstrate that both the mixed model and the multi-exponential model consistently outperform the single stretched exponential model, providing strong justification for our use of the mixed model in the main analysis.

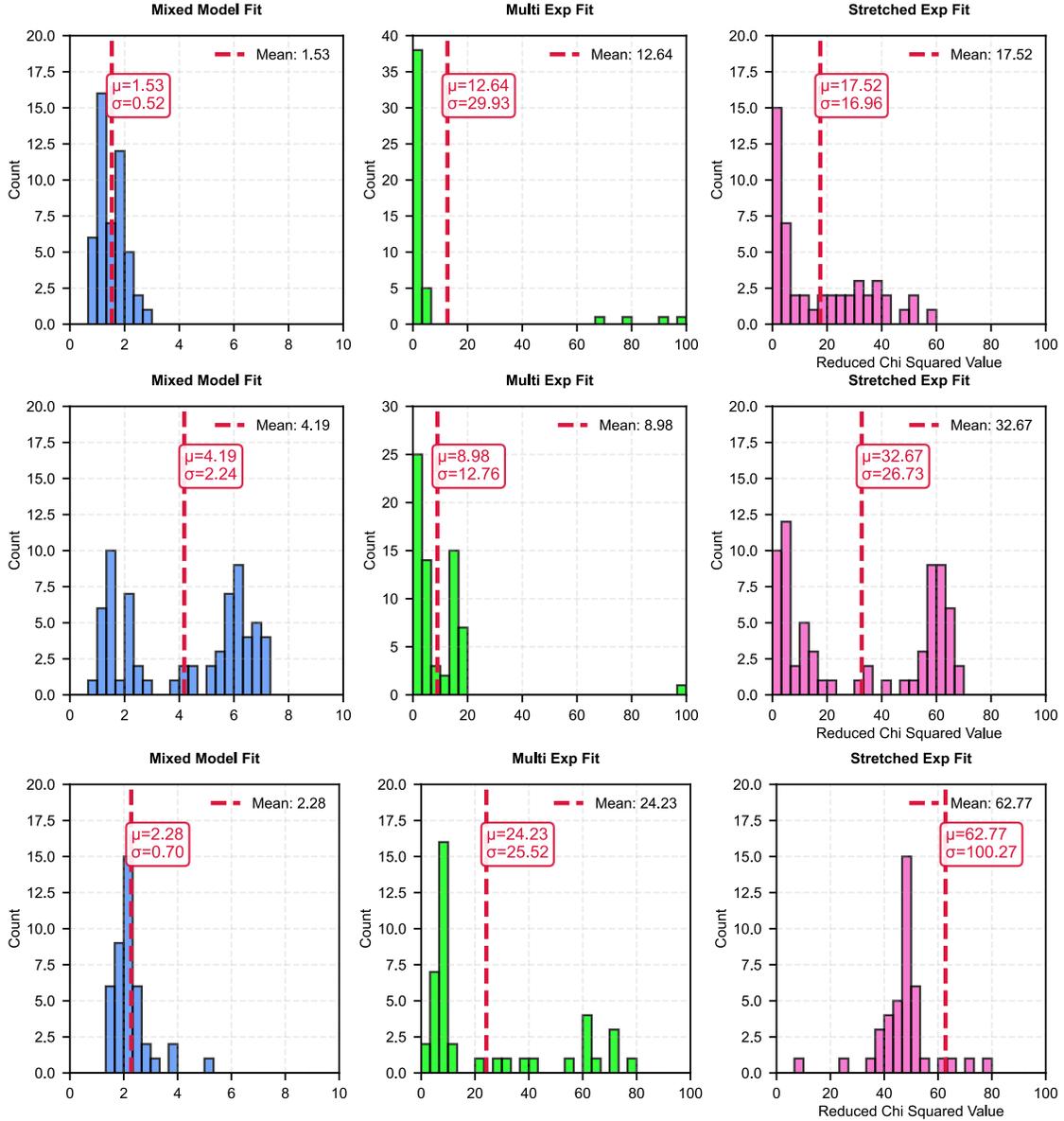

FIG. S7: **Comparison of reduced chi-squared values for sample Test-8 S6, H32-S1, and H33-S1.** Histograms show the distribution of the reduced chi-squared ($\chi^2_{red}$) values obtained from fitting the lifetime decay data with three different models: the mixed model (left), the multi-exponential model (center), and the stretched exponential model (right). The distribution for the mixed model is tightly centered near unity ($\mu = 1.53$), indicating a consistently robust fit. In contrast, the other models yield significantly larger average $\chi^2_{red}$ values and broader distributions.

**S6.6 Power dependence of effective decay time**

$$\frac{dn}{dt} = D\nabla n - \frac{1}{\tau_{\mathrm{sp}}}n - Bn^2 + Cn^3 \tag{S4}$$

One can now assume the decay dynamics can approximate with some effective decay time, and write

$$n(t) = n_0 e^{-t/\tau_{\mathrm{eff}}} \tag{S5}$$

Clearly, one can find the effective decay time as

$$\tau_{\mathrm{eff}} = -\frac{n_0}{dn(t)/dt|_{t=0}} \tag{S6}$$

Using this and the equation S4, and ignoring the contribution from diffusion term we can now express the effective decay time as

$$\tau_{\mathrm{eff}} = \frac{\tau_{\mathrm{sp}}}{1 + \tau_{\mathrm{sp}}Bn_0 + \tau_{\mathrm{sp}}Cn_0^2} \tag{S7}$$

Now since the intial exciton density is proportional to input power $n_0 = GP$ at low power limit, we can get effective power dependence for observed decay time

$$\tau_{\mathrm{eff}} = \frac{\tau_{\mathrm{sp}}}{1 + \tau_{\mathrm{sp}}BGP + \tau_{\mathrm{sp}}G^2CP^2} \tag{S8}$$

## S6.  LOW TEMPERATURE SPECTRA

Emission peak for WSe$_2$ fitted via Julia curve fitting function with Gaussian and Voigt distributions with propagating error statistics. Residues plotted as difference between raw spectra data and fitted curve amplitudes. Eight peaks between 1.55 eV and 1.74 eV were identified.

**S5.1 Power and temperature dependence of PL spectra**

Power-dependent spectra measured at 5 K from 30 mW to 360 mW. Temperature dependent data measured at 55 mW input power from 5 K up to 100 K. Peak amplitudes of each identified Gaussian and Voigt peaks and peak energies were plotted and fitted against the power law to match the distribution identifying as exciton, bi-exciton, or trion through coefficient $\alpha$.

**Lorentz model [1]:** The causal Lorentz model for emission/absorption spectra is,

$$f_{\mathrm{L}}(\omega; A_{\mathrm{L}}, \omega_0, \gamma) = A_{\mathrm{L}} \frac{\gamma\omega}{(\omega_0^2 - \omega^2)^2 + 4\gamma^2\omega^2}$$

Decomposing the Lorentzian into single poles with $\omega_0^2 = \Omega_{\mathrm{L}}^2 + \gamma^2$ gives [2],

$$f_{\mathrm{L}}(\omega; A_{\mathrm{L}}, \omega_0, \gamma) = f_{\mathrm{L}}(\omega; A_{\mathrm{L}}, \Omega_{\mathrm{L}}, \gamma) = \frac{A_{\mathrm{L}}}{2\Omega_{\mathrm{L}}} \left[ \frac{\gamma}{(\omega - \Omega_{\mathrm{L}})^2 + \gamma^2} - \frac{\gamma}{(\omega + \Omega_{\mathrm{L}})^2 + \gamma^2} \right],$$

It is a common practice to ignore the second (blue-marked) part in spectroscopic fits and use a non-causal, single-term version,

$$f_{\mathrm{L}}(\omega; A_{\mathrm{L}}, \Omega_{\mathrm{L}}, \gamma) = A_{\mathrm{L}} \frac{\gamma}{\gamma^2 + (\omega - \Omega_{\mathrm{L}})^2}$$

that also matches the scaled Cauchy-Lorentz probability density function (PDF).

**Causal Gaussian model (see e.g., [2–4]):**

$$f_{\mathrm{G}}(\omega; A_{\mathrm{G}}, \Omega_{\mathrm{G}}, \sigma) = \frac{A_{\mathrm{G}}\sqrt{\pi}}{2\sigma\sqrt{2}}\left[\mathrm{e}^{-\frac{(\omega-\Omega_{\mathrm{G}})^2}{2\sigma^2}} - \mathrm{e}^{-\frac{(\omega+\Omega_{\mathrm{G}})^2}{2\sigma^2}}\right]$$

We can also ignore the second (orange-marked) part in spectroscopic fits and use a non-causal, single-term version,

$$f_{\mathrm{G}}(\omega; A_{\mathrm{G}}, \Omega_{\mathrm{G}}, \sigma) = \frac{A_{\mathrm{G}}\sqrt{\pi}}{2\sigma\sqrt{2}}\mathrm{e}^{-\frac{(\omega-\Omega_{\mathrm{G}})^2}{2\sigma^2}}$$

which is a scaled Gaussian PDF.

**Causal Voigt (Gauss-Lorentz) model [2, 5]:**

$$f_{\mathrm{V}}(\omega; A_{\mathrm{V}}, \Omega_{\mathrm{V}}, \gamma) = \frac{A_{\mathrm{V}}\sqrt{\pi}}{2\sigma\Omega_{\mathrm{V}}\sqrt{2}}\Re\left[w\left(\frac{\omega+\iota\gamma-\Omega_{\mathrm{V}}}{\sigma\sqrt{2}}\right) - w\left(\frac{\omega+\iota\gamma+\Omega_{\mathrm{V}}}{\sigma\sqrt{2}}\right)\right]$$

Ignoring the last (purple-marked) term yields a simpler, single-term version,

$$f_{\mathrm{V}}(\omega; A_{\mathrm{V}}, \Omega_{\mathrm{V}}, \gamma) = \frac{A_{\mathrm{V}}\sqrt{\pi}}{2\sigma\Omega_{\mathrm{V}}\sqrt{2}}\Re\left[w\left(\frac{\omega+\iota\gamma-\Omega_{\mathrm{V}}}{\sigma\sqrt{2}}\right)\right]$$

The single-term lineshape $f_{\mathrm{V}}(\omega; ...)$ above is obtained as the convolution of the single-term Lorentzian and Gaussian functions $f_{\mathrm{V}}(\omega) = (f_{\mathrm{L}} * f_{\mathrm{G}})(\omega)$; $w(z)$ denotes the Faddeeva (Kramp) function of a complex argument

$$z = \frac{\omega - \Omega_V + i\gamma}{\sigma\sqrt{2\pi}},$$

which apparently inherits all the critical parameters, i.e., offset $\Omega_{\mathrm{V}} = \Omega_{\mathrm{G}} + \Omega_{\mathrm{L}}$ and broadening ($\gamma$ and $\sigma$), from both convolved functions. The Faddeeva function [6] is related to the scaled complex complementary error function as

$$w(z) = \exp\left(-z^2\right)\mathrm{erfc}(-iz) = \mathrm{erfcx}(-iz) = \exp\left(-z^2\right)\left[1 + \frac{2i}{\sqrt{\pi}}\int_0^z \mathrm{e}^{t^2}\mathrm{d}t\right]$$

An open-source C++ code for approximating the Faddeeva function (with wrappers for C, Matlab, GNU Octave, Python, R, Scilab, and Julia) was written by Steven Johnson [7]. Very efficient, few-term approximants for $w(z)$ are also available in [2, 4].

**Fitting with the Single-Term Gauss and Voigt functions:** The fitted $\Omega_{\mathrm{G,V}}$ are defined as the peak energy for the eight peaks that correspond to Raman spectra. The residues are taken as the difference between the fitted spectra and the raw spectra. The subsequent fitted parameters are fitted again into a power law. Both the Voigt and Gaussian models match the curves well with the same number of peaks from low to high input power. Several local maximum points in the residues arise as a result of negative counts in the spectrometer noise. Several fitting uncertainties arise from the interplay of overlapping peaks that are optimized by setting the fit tolerance in the code. Energies of each peak remain consistent in power-dependent spectra, with negligible changes arising from fitting errors. Peak energies blue shifts with increasing temperature, albeit with large fitting uncertainties and poor mapping of each peak's change as a function of temperature due to some peak disappearing at higher temperatures.

**Power law:**

$$I(P_{input}) = AP_{\mathrm{in}}^{\alpha}$$

The fitted Gaussian and Voigt peaks in temperature-dependent data were further fitted as part of the Varshni equation to obtain the binding energies. The binding energies and the amount of sample doping from the Au-assisted transfer process can be further extracted.

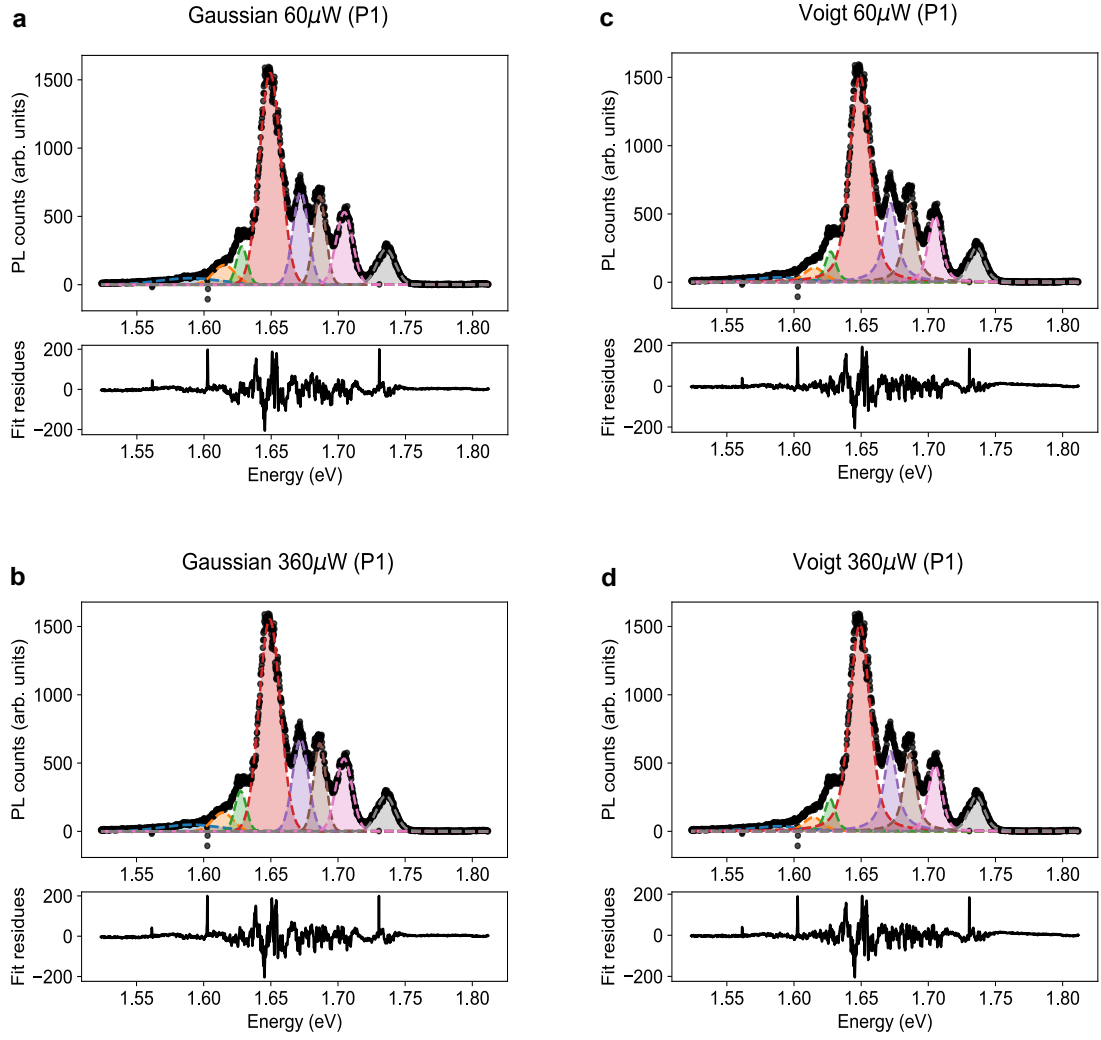

FIG. S8: **Comparison of Gaussian and Voigt peak fitting models for power-dependent photoluminescence spectra.** The emission spectrum of monolayer WSe$_2$ at 5 K is fit with a sum of eight peaks using either **a,b**, Gaussian or **c,d**, Voigt lineshape. The fits are shown for low (60 $\mu$W) and high (360 $\mu$W) excitation powers. For each panel, the raw data (black dots), the total fit (black line), and the individual fitted peaks (colored shaded areas) are shown. The lower panels display the fit residuals, calculated as the difference between the raw data and the total fit.

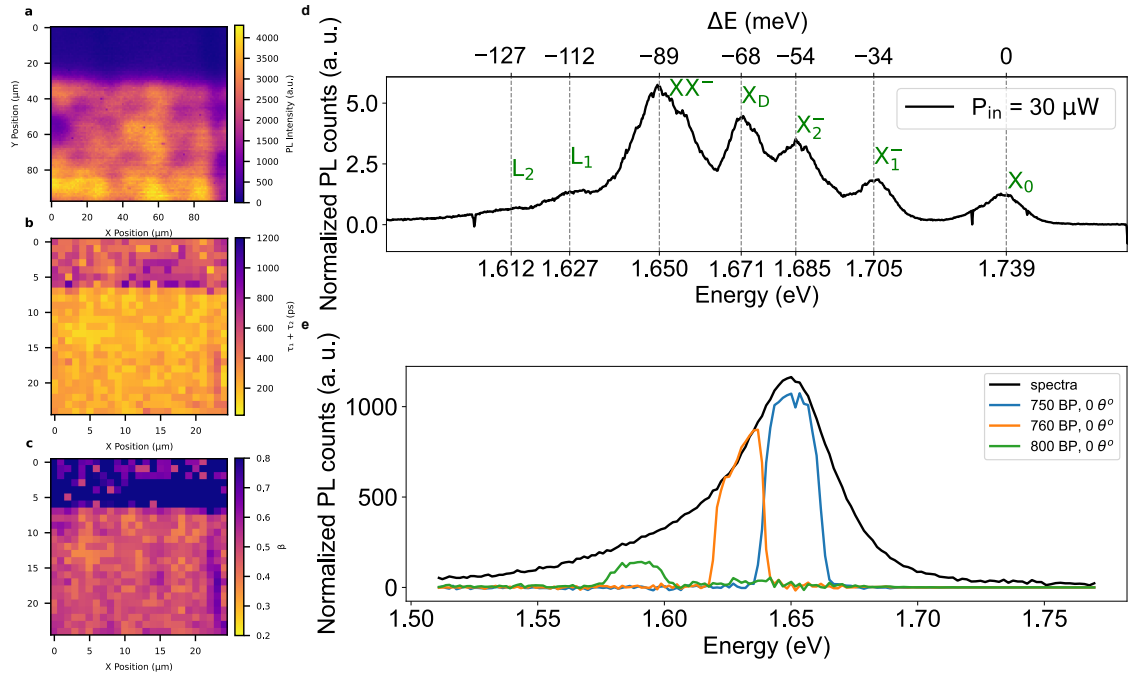

FIG. S9: **a**, Photoluminescence intensity scan. **b**, Map of the average lifetime ($\tau_{\mathrm{avg}}$). **c**, Map of the stretching exponent ($\beta$). **d**, PL spectrum with exciton peaks identified. **e**, PL spectrum (blue) of the sample (H33-S1) with transmission spectra of the bandpass filters used for energy-selective lifetime measurements: 750 nm (green, $0°$ tilt), 760 nm (orange, $0°$ tilt), and 800 nm (red) tilted by $25°$ to achieve a center wavelength of 780 nm.

## S7. ROOM TEMPERATURE OPTICAL SETUP

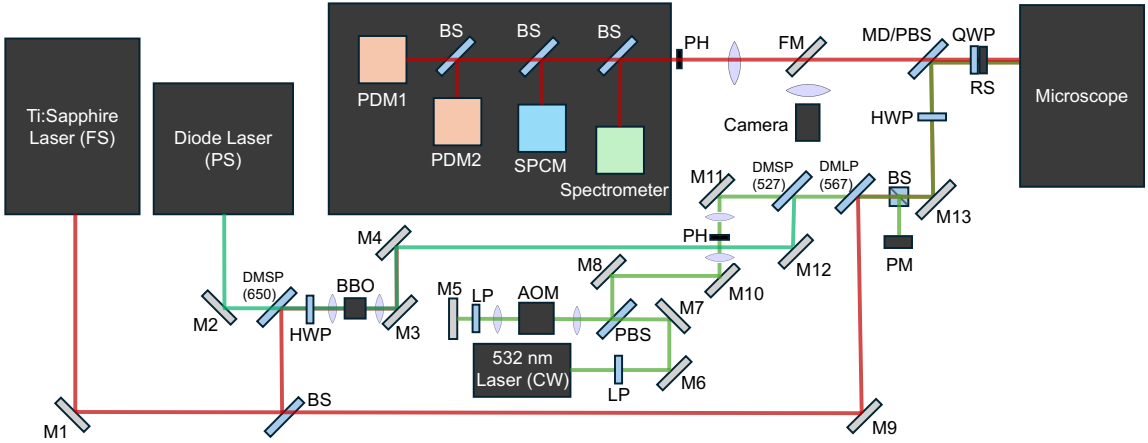

FIG. S10: **Schematic of the room-temperature optical setup.** The lasers include a Mai Tai HP DeepSee Ti:Sapphire laser (690–1040 nm, 150 fs pulse width, 80 MHz) and a continuous-wave 532 nm laser from RGB Photonics. A Becker & Hickl picosecond diode laser (515 nm, 40 ps, 80 MHz) is shown but was not used. A beta barium borate (BBO) crystal enables second harmonic generation (SHG) for tunable sources. Optical elements include silver mirrors (M1, M2, ...), acousto-optical modulator (AOM, Gooch & Housego 3350-199), linear polarizers (LP), half-wave plates (HWP), beam splitters (BS), polarizing beam splitters (PBS), a quarter-wave plate (QWP) and a 100 $\mu$m pinhole filter (PH). The QWP is mounted on a rotation stage (RS). Spectral filtering is done via dichroic mirrors: short-pass (DMSP), long-pass (DMLP), and main dichroic (MD). For resonant excitation at $\lambda = 750$ nm, the main dichroic was replaced with a broadband polarizing beam splitter. Photon counting was performed using SPADs (SPCM-AQRH, Excelitas; PDM series, Micro-Photon Devices). Spectra were collected using a QE65000 spectrometer (Ocean Insight).

Room-temperature measurements were conducted on a custom-built scanning confocal microscope, depicted in Fig. S10. The system is based on a Nikon Ti-U inverted microscope body. A Nikon LU Plan 100×/0.90 NA objective focuses the excitation laser and collects the resulting photoluminescence (PL). The sample is mounted on a three-axis piezo stage (Physik Instrumente, P-561 PIMars) controlled by an E-712 controller for precise positioning and raster scanning.

The optical path includes several key components. For continuous-wave (CW) excitation, a 532 nm laser (RGB Photonics) was used. For time-resolved measurements, a tunable Ti:Sapphire femtosecond laser (Spectra-Physics, Mai Tai HP; 690–1040 nm, 150 fs, 80 MHz) provided pulsed excitation, with a beta barium borate (BBO) crystal used for second harmonic generation where necessary.

The collected PL signal is directed through a main dichroic beamsplitter (Thorlabs, DMLP550L) and spectrally filtered by a 550 nm long-pass filter (Thorlabs, FEL0550) and a 750 ± 10 nm band-pass filter (Thorlabs, FBH750-10). In the case of spectrally filtered lifetime measurements (Main text, Fig. 4), the 750 BP was swapped for either the 550 LP, 760 BP, or 780 BP as indicated. The light is then focused through a 100 μm pinhole for spatial filtering. For CW and scanning measurements, photons are detected with a single-photon avalanche diode (SPAD; Excelitas, SPCM-AQRH; 69% quantum efficiency at 650 nm). For lifetime measurements, a high-resolution SPAD (Micro-Photon Devices, PDM series; 30 ps resolution, 35% quantum efficiency at 650 nm) is used in conjunction with a Becker & Hickl SPC-150N time-correlated single-photon counting (TCSPC) module. Emission spectra were acquired using a QE65000 spectrometer (Ocean Insight). All hardware was integrated and controlled via a custom LabVIEW interface.

## S8. EXTRACTION OF DOPING DENSITY

To determine the unintentional doping level in our monolayer WSe$_2$, we analyze the exciton and trion resonances observed in low-temperature photoluminescence (PL) spectra. The clear appearance of the negative trion ($X^-$) peak identifies the doping as $n$-type (electron doping). This is consistent with the fact that only electron doping stabilizes $X^-$ states in WSe$_2$ [8].

From the PL spectra, we extract the exciton–trion energy splitting

$$\Delta E = E_{X^0} - E_{X^-} \approx 34 \text{ meV},$$

with $E_{X^0} = 1.739$ eV and $E_{X^-} = 1.705$ eV. In the high-density regime, where the electron Fermi energy exceeds the trion binding energy, the splitting is approximately given by [9]

$$\Delta E \approx E_F + E_{\text{bind}}^{X^-}.$$

Taking the trion dissociation energy as $E_{\text{bind}}^{X^-} \approx 5\text{-}6$ meV ($\sim 60\text{-}70$ K), the measured splitting corresponds to a Fermi energy on the order of tens of meV. Using the 2D relation

$$E_F = \frac{\pi \hbar^2 n}{m_e^*},$$

with the conduction-band effective mass $m_e^* \approx 0.3 \, m_0$, this yields an electron density in the range

$$n_e \sim 10^{11}\text{–}10^{12} \text{ cm}^{-2}.$$